\documentclass[11pt,onecolumn,aps,superscriptaddress,
               amsmath,amssymb,nofootinbib]{revtex4}      
\usepackage{graphicx}

\usepackage{graphicx,epsfig}

\usepackage{dcolumn}
\usepackage{bm}

\newcommand{\Z}{{\mathbb Z}}

\begin{document}

\begin{flushright}
\baselineskip=12pt \normalsize
{ACT-06-07},
{MIFP-07-29}\\
\smallskip
\end{flushright}

\title{Towards Realistic Susy Spectra and Yukawa Textures from Intersecting Branes}

\author{Ching-Ming Chen}
\affiliation{George P. and Cynthia W. Mitchell Institute for
Fundamental Physics, Texas A\&M
University,\\ College Station, TX 77843, USA}
\author{Tianjun Li}
\affiliation{George P. and Cynthia W. Mitchell Institute for
Fundamental Physics, Texas A\&M
University,\\ College Station, TX 77843, USA}
\affiliation{Institute of Theoretical Physics, Chinese Academy of Sciences, Beijing 100080, China}
\author{V.E. Mayes}
\affiliation{George P. and Cynthia W. Mitchell Institute for
Fundamental Physics, Texas A\&M
University,\\ College Station, TX 77843, USA}
\author{D.V. Nanopoulos}
\affiliation{George P. and Cynthia W. Mitchell Institute for
Fundamental Physics, Texas A\&M
University,\\ College Station, TX 77843, USA}
\affiliation{Astroparticle Physics Group, Houston
Advanced Research Center (HARC),
Mitchell Campus,
Woodlands, TX~77381, USA; \\
Academy of Athens,
Division of Natural Sciences, 28~Panepistimiou Avenue, Athens 10679,
Greece}

\begin{abstract}
\begin{center}
{\bf ABSTRACT}
\end{center} 
We study the possible phenomenology of a
three-family Pati-Salam model constructed from intersecting D6-branes in
Type IIA string theory on the $\mathbf{T^6/(\Z_2\times \Z_2)}$
orientifold with some desirable semi-realistic features. In the model, tree-level 
gauge coupling unification is
achieved automatically at the string scale, and the gauge symmetry
may be broken to the Standard Model (SM) close to the string
scale. The small number of extra chiral exotic states in the model may be decoupled
via the Higgs mechanism and strong dynamics. We calculate the possible
supersymmetry breaking soft terms and the corresponding
low-energy supersymmetric particle spectra which may potentially
be tested at the Large Hadron Collider (LHC). We find that for the viable
regions of the parameter space the lightest CP-even
Higgs boson mass usually satisfies $m_H \leq 120$~GeV, and the observed
dark matter density may be generated.   
Finally, we find that it
is possible to obtain correct SM quark masses and mixings,
and the tau lepton mass at the unification scale.  Additionally, 
neutrino masses and mixings
may be generated via the seesaw mechanism.  
Mechanisms to stabilize the open and closed-string moduli,
which are necessary for the model to be truly viable and to make
definite predictions are discussed.  
\end{abstract}

\maketitle

\newpage
\section{Introduction}
Although string theory has long teased us with her power to
encompass all known physical phenomena in a complete mathematical
structure, an actual worked out example is  still lacking. Indeed,
the major problem of string phenomenology is to construct {\it at
least one} realistic model with all moduli stabilized, which
completely describes known particle physics as well as potentially
being predictive of unknown phenomena. With the dawn of the 
Large Hadron Collider (LHC)
era, new discoveries will hopefully be upon us. In particular,
supersymmetry is expected to be found as well as the Higgs states
required to break the electroweak symmetry. Therefore, it is
highly desirable to have complete, concrete models derived from string theory
which are able to make predictions for the superpartner spectra,
as well as describing currently known particle physics.

In the old days of string phenomenology, model builders were
primarily focused on weakly coupled heterotic string theory.
However, with the advent of the second string revolution,
D-branes~\cite{JPEW} have created new interest in Type I and II
compactifications.  In particular, Type IIA orientifolds with
intersecting D6-branes, where the chiral fermions arise at
the intersections of D6-branes in
the internal space~\cite{bdl}, with T-dual Type IIB description in
terms of magnetized D-branes~\cite{bachas}, have shown great
promise during the last few years.  Indeed, intersecting D-brane configurations 
provide promising setups
which may accommodate semi-realistic features of low-energy physics. Given this,
it is an interesting question to see how far one can get from a particular
string compactification to reproducing the finer details of the Standard 
Model as a low-energy effective field theory.  

In order to construct globally consistent vacua with intersecting D-branes, 
conditions must be imposed which 
strongly constrain the models.  In particular, all Ramond-Ramond (RR) tadpoles must
be cancelled and 
K-theory \cite{Witten9810188}
conditions for cancelling the nontrivial $Z_2$ anomally also must
be imposed. 
Despite the clear benefits of supersymmetry, there have been many
three-family standard-like models and Grand Unified Theories (GUT)
constructed on Type IIA
orientifolds~\cite{Blumenhagen:2000wh, Angelantonj:2000hi,
Blumenhagen:2005mu} which are not supersymmetric.  Although these
models are globally consistent, they are generally plagued by the
gauge hierarchy problem and vacuum instability which arises from
uncancelled
Neveu-Schwarz-Neveu-Schwarz (NSNS) tadpoles.  Later,
semi-realistic supersymmetric Standard-like, Pati-Salam, unflipped
$SU(5)$ as well as flipped $SU(5)$ models in Type IIA theory on
$\mathbf{T^6/(\Z_2\times \Z_2)}$~\cite{CSU1, CSU2, Cvetic:2002pj,
CP, CLL, Cvetic:2004nk, Chen:2005ab, Chen:2005mj} and
$\mathbf{T^6/(\Z_2\times \Z_2')}$~\cite{Dudas:2005jx, Blumenhagen:2005tn,
Chen:2006sd} orientifolds were eventually constructed, and some of
their phenomenological consequences studied~\cite{CLS1, CLW}.
Other supersymmetric constructions in Type IIA theory on different
orientifold backgrounds have also been
discussed~\cite{ListSUSYOthers}.  Nonperturbative D-instanton effects
have also been receiving much attention of late, and may play an important
role~\cite{Blumenhagen:2006xt}~\cite{Ibanez:2006da}~\cite{Cvetic:2007ku}~\cite{Ibanez:2007rs}.

In addition to satisfying the above consistency conditions, all open and closed-string
moduli must be stabilized in order to obtain an actual vacuum.  Unstabilized 
moduli are manifest in the low-energy theory as massless scalar fields, which are clearly
in conflict with observations.  
Given a concrete string model, the low-energy observables such as particle
couplings and resulting masses are functions of the open and closed string
moduli. In a fully realistic model, these moduli must therefore be stabilized
and given sufficiently large masses to meet the astrophysical/cosmological and collider
physics constraints on additional scalar fields.  Although satisfying the conditions for
$\mathcal{N}=1$ supersymmetry in Type IIA (IIB) fixes the complex structure (K\a"ahler) moduli in these
models, the K\a"ahler (complex structure) and open-string moduli generally remain
unfixed.  
To stabilize some of these moduli, supergravity
three-form fluxes \cite{Kachru:Blumen} and geometric fluxes
\cite{Grimm:Zwirner} were introduced and  flux models on Type II
orientifolds have been constructed~\cite{Cascales:2003zp, MS, CL,
Cvetic:2005bn, Kumar:2005hf, Chen:2005cf, Blumenhagen:2006ci, Camara:2005dc,
Chen:2006gd, Chen:2006ip}.  Models where the D-branes wrap rigid cycles, thus
freezing the open-string moduli have also been studied~\cite{Dudas:2005jx, Blumenhagen:2005tn,
Chen:2006sd}.

Despite substantial progress, there have been other roadblocks in
constructing phenomenologically realistic intersecting D-brane models, besides the
usual problem of moduli stabilization.  Unlike
heterotic models, the gauge couplings are not automatically
unified.  Additionally, there has been a rank one problem in the
Standard Model (SM) fermion Yukawa matrices, preventing the
generation of mass for the first two generations of quarks and leptons.
For the case of toroidal orientifold compactifications, this can be
traced to the fact that not all of the Standard Model fermions
are localized at intersections on the same torus.
However, one example of an intersecting D6-brane model on Type IIA
$\mathbf{T^6/(\Z_2\times \Z_2)}$ orientifold has recently been
discovered where these problems may be
solved~\cite{CLL,Chen:2006gd}.  Thus, this particular model may be
a step forward to obtaining realistic phenomenology from string
theory. Indeed, as we recently discussed ~\cite{Chen:2007px}, it
is possible within the moduli space of this model to obtain correct 
quark mass matrices
and mixings, the tau lepton mass, and to generate naturally small
neutrino masses via the seesaw mechanism. Furthermore, it is
possible to generically study the soft supersymmetry breaking
terms, from which can be calculated the supersymmetric partner
spectra, the Higgs masses, and the resulting neutralino relic
density.

This paper is organized as follows. First, we will briefly review
the intersecting D6-brane model on Type IIA
$\mathbf{T^6/(\Z_2\times \Z_2)}$ orientifold which we are studying
and discuss its basic features.  We then discuss the low-energy effective
action, and show that the tree-level 
gauge couplings are unified near the string scale.  
We also find that the hidden sector gauge groups 
will become confining at a high energy scale, thus decoupling chiral 
exotics present in the model. Next, we study the possible 
low-energy superpartner spectra which may arise. 
We also calculate the Yukawa
couplings for quarks and leptons in this model, and show that we may
obtain the correct quark masses and mixings and the tau lepton mass for
specific choices of the open and closed string-moduli VEVs.  We should
emphasize that for the present
work, we will not focus on the moduli stabilization problem, as our goal is
only to explore the possible phenemological characteristics of the model. 
However, we do comment on
this issue and discuss how it may
potentially be solved for this model. We also should note that models with an
equivalent observable sector have been constructed in Type IIA and Type IIB theory
as Ads and Minkowski flux vacua~\cite{Chen:2006gd, Chen:2007af}, so that
the issue of closed-string moduli stablization has already
been addressed to some extent.

\section{A D-brane Model with Desirable Semi-realistic Features}

In recent years, intersecting D-brane models have provided
an exciting approach towards constructing semi-realistic vacua. 
To summarize, 
D6 branes (in Type IIA) fill three-dimensional
Minkowski space and wrap 3-cycles in the compactified manifold, with a stack of
$N$ branes having a gauge group $U(N)$ (or $U(N/2)$ in the case
 of $T^6/(\Z_2 \times \Z_2)$) in its world volume. The 3-cycles 
wrapped by the D-branes will in general intersect multiple times in the
internal space, resulting in 
a chiral fermion in the bifundamental representation
localized at the intersection between different stacks. The multiplicity
of such fermions is then given by the number of times the 3-cycles intersect.
Due to orientifolding,
for every stack of D6-branes we must also introduce its orientifold
images.  Thus, the D6-branes may also have intersections with the images of other stacks,
also resulting in fermions in bifundamental representations.
Each stack may also intersect its own images, resulting in chiral fermions 
in the symmetric and antisymmetric representations.  The different
types of representations that may be obtained for each type of intersection
and their multiplicities are shown in Table~\ref{spectrum}.
In addition, there are constraints that must be satisfied for the 
consistency of the model, namely the requirement for Ramond-Ramond
tadpole cancellation and to have a sprectrum with $\mathcal{N}=1$
supersymmetry.

Intersecting D-brane configurations provide promising setups
which may accommodate semi-realistic features of low-energy physics. Given this,
it is an interesting question to see how far one can get from a particular
string compactification to reproducing the finer details of the Standard 
Model as a low-energy effective field theory.  
There have been many consistent models studied, but only a small number
have the proper structures to produce an acceptable phenomenology.  
A good candidate for a
realistic model which may possess the proper structures was discussed in
\cite{CLL,Chen:2006gd,Chen:2007px} in Type IIA theory
on the $\mathbf{T^6/(\Z_2\times \Z_2)}$ orientifold. This background has been extensively
studied and we refer the reader to~\cite{CSU1,CSU2} for reviews of
the basic model building rules. We present the D6-brane
configurations and intersection numbers of this model in
Table~\ref{MI-Numbers}, and the resulting spectrum which is
essentially that of a three-family Pati-Salam in
Table~\ref{Spectrum}~\cite{CLL,Chen:2006gd}. We put the $a'$, $b$,
and $c$ stacks of D6-branes on the top of each other on the third
two torus, and as a result there are additional vector-like
particles from $N=2$ subsectors.

\begin{table}[ht]
\caption{General spectrum for intersecting D6-branes at generic
angles, where $I_{aa'}=-2^{3-k}\prod_{i=1}^3(n_a^il_a^i)$, and
$I_{aO6}=2^{3-k}(-l_a^1l_a^2l_a^3
+l_a^1n_a^2n_a^3+n_a^1l_a^2n_a^3+n_a^1n_a^2l_a^3)$. Moreover,
${\cal M}$ is the multiplicity, and $a_S$ and $a_A$ denote
 the symmetric and anti-symmetric representations of
$U(N_a/2)$, respectively.}
\renewcommand{\arraystretch}{1.4}
\begin{center}
\begin{tabular}{|c|c|}
\hline {\bf Sector} & \phantom{more space inside this box}{\bf
Representation}
\phantom{more space inside this box} \\
\hline\hline
$aa$   & $U(N_a/2)$ vector multiplet  and 3 adjoint chiral multiplets  \\
\hline $ab+ba$   & $ {\cal M}(\frac{N_a}{2},
\frac{\overline{N_b}}{2})=
I_{ab}=2^{-k}\prod_{i=1}^3(n_a^il_b^i-n_b^il_a^i)$ \\
\hline $ab'+b'a$ & $ {\cal M}(\frac{N_a}{2},
\frac{N_b}{2})=I_{ab'}=-2^{-k}\prod_{i=1}^3(n_{a}^il_b^i+n_b^il_a^i)$ \\
\hline $aa'+a'a$ &  ${\cal M} (a_S)= \frac 12 (I_{aa'} - \frac 12
I_{aO6})$~;~~ ${\cal M} (a_A)=
\frac 12 (I_{aa'} + \frac 12 I_{aO6}) $ \\
\hline
\end{tabular}
\end{center}
\label{spectrum}
\end{table}

\begin{figure}[h]
\begin{center}
\includegraphics[width=.9\textwidth,angle=0]{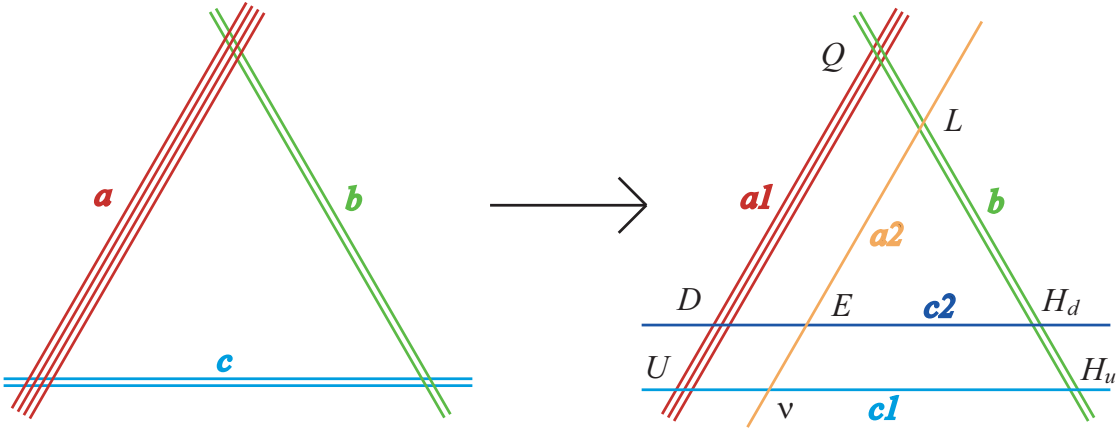}
\caption{Breaking of the effective gauge symmetry $SU(4)\times SU(2)_L\times SU(2)_R$ down to
$SU(3)_C\times U(2)_L\times U(1)_{I3R}\times U(1)_{B-L}$ via brane splitting. 
This process corresponds to giving a VEV to adjoint scalars, which arise as open-string moduli associated with the positions of stacks $a$ and $c$ in the internal space.} \label{brnsplit}
\end{center}
\end{figure}

The anomalies from three global $U(1)$s of $U(4)_C$, $U(2)_L$ and
$U(2)_R$ are cancelled by the Green-Schwarz mechanism, and the
gauge fields of these $U(1)$s obtain masses via the linear
$B\wedge F$ couplings. Thus, the effective gauge symmetry is
$SU(4)_C\times SU(2)_L\times SU(2)_R$. In order to break the gauge
symmetry, on the first torus, we split the $a$ stack of D6-branes
into $a_1$ and $a_2$ stacks with 6 and 2 D6-branes, respectively,
and split the $c$ stack of D6-branes into $c_1$ and $c_2$ stacks
with two D6-branes for each one, as shown in Figure
\ref{brnsplit}. In this way, the gauge symmetry is further broken
to $ SU(3)_C\times SU(2)_L\times U(1)_{I_{3R}}\times U(1)_{B-L}$.
Moreover, the $U(1)_{I_{3R}}\times U(1)_{B-L}$ gauge symmetry may
be broken to $U(1)_Y$ by giving vacuum expectation values (VEVs)
to the vector-like particles with the quantum numbers $({\bf { 1},
1, 1/2, -1})$ and $({\bf { 1}, 1, -1/2, 1})$ under the
$SU(3)_C\times SU(2)_L\times U(1)_{I_{3R}} \times U(1)_{B-L} $
gauge symmetry from $a_2 c_1'$
intersections~\cite{CLL,Chen:2006gd}.

\begin{table}[t]
\footnotesize
\renewcommand{\arraystretch}{1.0}
\caption{D6-brane configurations and intersection numbers for
the model on Type IIA $\mathbf{T}^6 / \Z_2 \times \Z_2$
orientifold. The complete gauge symmetry is $[U(4)_C \times U(2)_L
\times U(2)_R]_{\rm observable}\times [ USp(2)^4]_{\rm hidden}$, the SM
fermions and Higgs fields arise from the intersections on the
first two-torus, and the complex structure parameters are
$2\chi_1=6\chi_2=3\chi_3 =6$.}
\label{MI-Numbers}
\begin{center}
\begin{tabular}{|c||c|c||c|c|c|c|c|c|c|c|c|c|}
\hline
& \multicolumn{12}{c|}{$U(4)_C\times U(2)_L\times U(2)_R\times USp(2)^4$}\\
\hline \hline  & $N$ & $(n^1,l^1)\times (n^2,l^2)\times

(n^3,l^3)$ & $n_{S}$& $n_{A}$ & $b$ & $b'$ & $c$ & $c'$& 1 & 2 & 3 & 4 \\

\hline

    $a$&  8& $(0,-1)\times (1,1)\times (1,1)$ & 0 & 0  & 3 & 0 & -3 & 0 & 1 & -1 & 0 & 0\\

    $b$&  4& $(3,1)\times (1,0)\times (1,-1)$ & 2 & -2  & - & - & 0 & 0 & 0 & 1 & 0 & -3 \\

    $c$&  4& $(3,-1)\times (0,1)\times (1,-1)$ & -2 & 2  & - & - & - & - & -1 & 0 & 3 & 0\\

\hline

    1&   2& $(1,0)\times (1,0)\times (2,0)$ & \multicolumn{10}{c|}{$\chi_1=3,~
\chi_2=1,~\chi_3=2$}\\

    2&   2& $(1,0)\times (0,-1)\times (0,2)$ & \multicolumn{10}{c|}{$\beta^g_1=-3,~
\beta^g_2=-3$}\\

    3&   2& $(0,-1)\times (1,0)\times (0,2)$& \multicolumn{10}{c|}{$\beta^g_3=-3,~
\beta^g_4=-3$}\\

    4&   2& $(0,-1)\times (0,1)\times (2,0)$ & \multicolumn{10}{c|}{}\\

\hline

\end{tabular}

\end{center}

\end{table}

\begin{table}
[htb] \footnotesize
\renewcommand{\arraystretch}{1.0}
\caption{The chiral and vector-like superfields,
 and their quantum numbers
under the gauge symmetry $SU(4)_C\times SU(2)_L\times SU(2)_R
\times USp(2)_1 \times USp(2)_2 \times USp(2)_3 \times USp(2)_4$.}
\label{Spectrum}
\begin{center}
\begin{tabular}{|c||c||c|c|c||c|c|c|}\hline
 & Quantum Number
& $Q_4$ & $Q_{2L}$ & $Q_{2R}$  & Field \\
\hline\hline
$ab$ & $3 \times (4,\overline{2},1,1,1,1,1)$ & 1 & -1 & 0  & $F_L(Q_L, L_L)$\\
$ac$ & $3\times (\overline{4},1,2,1,1,1,1)$ & -1 & 0 & $1$   & $F_R(Q_R, L_R)$\\
$a1$ & $1\times (4,1,1,2,1,1,1)$ & $1$ & 0 & 0  & $X_{a1}$ \\
$a2$ & $1\times (\overline{4},1,1,1,2,1,1)$ & -1 & 0 & 0   & $X_{a2}$ \\
$b2$ & $1\times(1,2,1,1,2,1,1)$ & 0 & 1 & 0    & $X_{b2}$ \\
$b4$ & $3\times(1,\overline{2},1,1,1,1,2)$ & 0 & -1 & 0    & $X_{b4}^i$ \\
$c1$ & $1\times(1,1,\overline{2},2,1,1,1)$ & 0 & 0 & -1    & $X_{c1}$ \\
$c3$ & $3\times(1,1,2,1,1,2,1)$ & 0 & 0 & 1   &  $X_{c3}^i$ \\
$b_{S}$ & $2\times(1,3,1,1,1,1,1)$ & 0 & 2 & 0   &  $T_L^i$ \\
$b_{A}$ & $2\times(1,\overline{1},1,1,1,1,1)$ & 0 & -2 & 0   & $S_L^i$ \\
$c_{S}$ & $2\times(1,1,\overline{3},1,1,1,1)$ & 0 & 0 & -2   & $T_R^i$  \\
$c_{A}$ & $2\times(1,1,1,1,1,1,1)$ & 0 & 0 & 2   & $S_R^i$ \\
\hline\hline
$ab'$ & $3 \times (4,2,1,1,1,1,1)$ & 1 & 1 & 0  & \\
& $3 \times (\overline{4},\overline{2},1,1,1,1,1)$ & -1 & -1 & 0  & \\
\hline
$ac'$ & $3 \times (4,1,2,1,1,1,1)$ & 1 &  & 1  & $\Phi_i$ \\
& $3 \times (\overline{4}, 1, \overline{2},1,1,1,1)$ & -1 & 0 & -1  &
$\overline{\Phi}_i$\\
\hline
$bc$ & $6 \times (1,2,\overline{2},1,1,1,1)$ & 0 & 1 & -1   & $H_u^i$, $H_d^i$\\
& $6 \times (1,\overline{2},2,1,1,1,1)$ & 0 & -1 & 1   & \\
\hline
\end{tabular}
\end{center}
\end{table}

Since the gauge couplings in the Minimal Supersymmetric Standard
Model (MSSM) are unified at the GUT scale $\sim2.4\times 10^{16}$
GeV, the additional exotic particles present in the model must
necessarily become superheavy. To accomplish this it is first
assumed that the $USp(2)_1$ and $USp(2)_2$ stacks of D6-branes lie
on the top of each other on the first torus, so we have two pairs
of  vector-like particles $X_{12}^i$ with $USp(2)_1\times
USp(2)_2$ quantum numbers $(2,2)$. These particles can break
$USp(2)_1\times USp(2)_2$ down to the diagonal $USp(2)_{D12}$ near
the string scale by obtaining VEVs, and then states arising from intersections $a1$
and $a2$ may obtain vector-like masses close to the string scale
from superpotential terms of the form
\begin{equation}
W \subset X_{a1}~X_{a2}~X_{12}^i,
\end{equation}
where we neglect the couplings of order one.
Moreover,
we assume that the $T_R^i$ and $S_R^i$ obtain VEVs near the string
scale, and their VEVs satisfy the D-flatness of $U(1)_R$. 
We also assume that there exist various
suitable high-dimensional operators in the effective theory,
and thus the adjoint chiral superfields may obtain GUT-scale
masses via these operators. With
$T_R^i$ and $S_R^i$, we can give GUT-scale masses to the
particles from the intersections $c1$, $c3$, and $c_S$ via
the supepotential:
\begin{equation}
W \subset S_R^i X_{c1}~X_{c1} + T_{R}^i X_{c3}^j~X_{c3}^k + \frac{1}{M_{Pl}} S_R^i S_R^j T_R^k T_R^l.
\end{equation}
The beta function for $USp(2)_{D12}$ is $-4$ and the gauge coupling for $USp(2)_{D12}$ will become
strongly coupled around $5 \times 10^{12}$~GeV, and then we can give $5 \times 10^{12}$~GeV scale
VEVs to $S_L^i$ and preserve the D-flatness of $U(1)_L$.  
The remaining states 
may also obtain intermediate scale masses via
the operators
\begin{equation}
W \subset X_{b2}~X_{b2} S_l^i + \frac{X_{b2}~X_{b2}}{M_X} X_{b4}~X_{b4}.
\end{equation}
To have one pair
of light Higgs doublets, it is necessary to fine-tune the mixing
parameters of the Higgs doublets. In particular, the $\mu$ term
and the right-handed neutrino masses may be generated via the
following high-dimensional operators
\begin{eqnarray}
W \supset &&{{y^{ijkl}_{\mu}} \over {M_{\rm St}}} S_L^i S_R^j
H_u^k H_d^l + {{y^{mnkl}_{Nij}}\over {M^3_{\rm St}}} T_R^{m}
T_R^{n} \Phi_i \Phi_j  F_R^k  F_R^l ~,~\,
\label{eqn:HiggsSup}
\end{eqnarray}
where $y^{ijkl}_{\mu}$ and $y^{mnkl}_{Nij}$ are Yukawa couplings,
and $M_{\rm St}$ is the string scale. Thus, the $\mu$ term is TeV
scale and the right-handed neutrino masses can be in the range
$10^{10-14}$ GeV for $y^{ijkl}_{\mu} \sim 1$ and $y^{mnkl}_{Nij}
\sim 10^{(-7)-(-3)}$.

\section{The $\mathcal{N}=1$ Low-energy Effective Action}

In building a concrete string model which may be testable, 
it is not enough to simply
reproduce the matter and gauge symmetry of the known low-energy
particle states in the Standard Model.  It is also necessary to
make predictions regarding the superpartner spectra and Higgs
masses.  If supersymmetry exists as expected and is softy broken,
then it is possible to calculate the soft SUSY breaking terms,
which determine the low energy sparticle spectra.  Furthermore,
if the neutralino is the lightest supersymmetric particle (LSP),
then it is expected to make up a large fraction of the observed
dark matter density, $0.0945~<\Omega h^2~<0.1287$ at
$2\sigma$~\cite{Bennett:2003bz, Spergel:2003cb}, and this is
calculable from the soft terms.  Ideally, one would also like
to be able to calculate the Yukawa couplings for the known quarks
and leptons, and be able to reproduce their masses and mixings.  

To discuss the low-energy phenomenology we start from the low-energy
effective action.  From the effective scalar potential it is
possible to study the stability ~\cite{Blumenhagen:2001te}, the
tree-level gauge couplings \cite{CLS1, Shiu:1998pa,
Cremades:2002te}, gauge threshold corrections \cite{Lust:2003ky},
and gauge coupling unification \cite{Antoniadis:Blumen}.  The
effective Yukawa couplings \cite{Cremades:2003qj, Cvetic:2003ch},
matter field K\"ahler metric and soft-SUSY breaking terms have
also been investigated \cite{Kors:2003wf}.  A more detailed
discussion of the K\"ahler metric and string scattering of gauge,
matter, and moduli fields has been performed in
\cite{Lust:2004cx}. Although turning on Type IIB 3-form fluxes can
break supersymmetry from the closed string sector
~\cite{Cascales:2003zp, MS, CL, Cvetic:2005bn, Kumar:2005hf,
Chen:2005cf}, there are additional terms in the superpotential
generated by the fluxes and there is currently no satisfactory
model which incorporates this. Thus, we do not consider this option
in the present work.  In principle, it should be possible to
specify the exact mechanism by which supersymmetry is broken, and
thus to make very specific predictions.  However, for the present
work, we will adopt a parametrization of the SUSY breaking so that
we can study it generically.

The $\mathcal{N}=1$ supergravity action depends upon three
functions, the holomorphic gauge kinetic function, $f$, K\a"ahler
potential $K$, and the superpotential $W$.  Each of these will in
turn depend upon the moduli fields which describe the background
upon which the model is constructed. The holomorphic gauge kinetic
function for a D6-brane wrapping a calibrated three-cyce is given
by~\cite{Blumenhagen:2006ci}
\begin{equation}
f_P = \frac{1}{2\pi \ell_s^3}\left[e^{-\phi}\int_{\Pi_P} \mbox{Re}(e^{-i\theta_P}\Omega_3)-i\int_{\Pi_P}C_3\right].
\end{equation}
In terms of the three-cycle wrapped by the stack of branes, we have
\begin{equation}
\int_{\Pi_a}\Omega_3 = \frac{1}{4}\prod_{i=1}^3(n_a^iR_1^i + 2^{-\beta_i}il_a^iR_2^i).
\end{equation}
from which it follows that
\begin{eqnarray}
f_P &=&
\frac{1}{4\kappa_P}(n_P^1\,n_P^2\,n_P^3\,s-\frac{n_P^1\,l_P^2\,l_P^3\,u^1}{2^{(\beta_2+\beta_3)}}-\frac{n_P^2\,l_P^1\,l_P^3\,u^2}{2^{(\beta_1+\beta_3)}}-
\frac{n_P^3\,l_P^1\,l_P^2\,u^3}{2^{(\beta_1+\beta_2)}}),
\label{kingauagefun}
\end{eqnarray}
where $\kappa_P = 1$ for $SU(N_P)$ and $\kappa_P = 2$ for
$USp(2N_P)$ or $SO(2N_P)$ gauge groups and where we use the $s$ and
$u$ moduli in the supergravity basis.  In the string theory basis,
we have the dilaton $S$, three K\"ahler moduli $T^i$, and three
complex structure moduli $U^i$~\cite{Lust:2004cx}. These are related to the
corresponding moduli in the supergravity basis by
\begin{eqnarray}
\mathrm{Re}\,(s)& =&
\frac{e^{-{\phi}_4}}{2\pi}\,\left(\frac{\sqrt{\mathrm{Im}\,U^{1}\,
\mathrm{Im}\,U^{2}\,\mathrm{Im}\,U^3}}{|U^1U^2U^3|}\right)
\nonumber \\
\mathrm{Re}\,(u^j)& =&
\frac{e^{-{\phi}_4}}{2\pi}\left(\sqrt{\frac{\mathrm{Im}\,U^{j}}
{\mathrm{Im}\,U^{k}\,\mathrm{Im}\,U^l}}\right)\;
\left|\frac{U^k\,U^l}{U^j}\right| \qquad (j,k,l)=(\overline{1,2,3})
\nonumber \\
\mathrm{Re}(t^j)&=&\frac{i\alpha'}{T^j} \label{idb:eq:moduli}
\end{eqnarray}
and $\phi_4$ is the four-dimensional dilaton.
To second order in the string matter fields, the K\a"ahler potential is given by
\begin{eqnarray}
K(M,\bar{M},C,\bar{C}) = \hat{K}(M,\bar{M}) + \sum_{\mbox{untwisted}~i,j} \tilde{K}_{C_i \bar{C}_j}(M,\bar{M})C_i \bar{C}_j + \\ \nonumber \sum_{\mbox{twisted}~\theta} \tilde{K}_{C_{\theta} \bar{C}_{\theta}}(M,\bar{M})C_{\theta}\bar{C}_\theta.
\end{eqnarray}
The untwisted moduli $C_i$, $\bar{C}_j$ are light, non-chiral
scalars from the field theory point of view, associated with the
D-brane positions and Wilson lines.  These fields are not observed
in the MSSM, and if they were present in the low energy spectra
may disrupt the gauge coupling unification.  Clearly, these
fields must get a large mass through some mechanism.  One way to
accomplish this is to require the D-branes to wrap rigid cycles,
which freezes the open string moduli~\cite{Blumenhagen:2005tn}.
However, there are no rigid cycles available on $T^6/(\Z_2 \times \Z_2)$
without discrete torsion, 
thus we
will assume that the open-string moduli become massive via high-dimensional
operators.

For twisted moduli arising from strings stretching between stacks
$P$ and $Q$, we have $\sum_j\theta^j_{PQ}=0$, where $\theta^j_{PQ} =
\theta^j_Q - \theta^j_P$ is the angle between the cycles wrapped
by the stacks of branes $P$ and $Q$ on the $j^{th}$ torus
respectively. Then, for the K\a"ahler metric in Type IIA theory we find
the following two cases:

\begin{itemize}

\item $\theta^j_{PQ}<0$, $\theta^k_{PQ}>0$, $\theta^l_{PQ}>0$

\begin{eqnarray}
\tilde{K}_{PQ} &=& e^{\phi_4} e^{\gamma_E (2-\sum_{j = 1}^3
\theta^j_{PQ}) }
\sqrt{\frac{\Gamma(\theta^j_{PQ})}{\Gamma(1+\theta^j_{PQ})}}
\sqrt{\frac{\Gamma(1-\theta^k_{PQ})}{\Gamma(\theta^k_{PQ})}}
\sqrt{\frac{\Gamma(1-\theta^l_{PQ})}{\Gamma(\theta^l_{PQ})}}
\nonumber \\ && (t^j + \bar{t}^j)^{\theta^j_{PQ}} (t^k +
\bar{t}^k)^{-1+\theta^k_{PQ}} (t^l +
\bar{t}^l)^{-1+\theta^l_{PQ}}.
\end{eqnarray}

\item $\theta^j_{PQ}<0$, $\theta^k_{PQ}<0$, $\theta^l_{PQ}>0$

\begin{eqnarray}
\tilde{K}_{PQ} &=& e^{\phi_4} e^{\gamma_E (2+\sum_{j = 1}^3
\theta^j_{PQ}) }
\sqrt{\frac{\Gamma(1+\theta^j_{PQ})}{\Gamma(-\theta^j_{PQ})}}
\sqrt{\frac{\Gamma(1+\theta^k_{PQ})}{\Gamma(-\theta^k_{PQ})}}
\sqrt{\frac{\Gamma(\theta^l_{PQ})}{\Gamma(1-\theta^l_{PQ})}}
\nonumber \\ && (t^j + \bar{t}^j)^{-1-\theta^j_{PQ}} (t^k +
\bar{t}^k)^{-1-\theta^k_{PQ}} (t^l + \bar{t}^l)^{-\theta^l_{PQ}}.
\end{eqnarray}

\end{itemize}

For branes which are parallel on at least one torus, giving rise
to non-chiral matter in bifundamental representations (for example,
the Higgs doublets), the K\a"ahler metric is
\begin{equation}
\hat{K}=((s+\bar{s})(t^1+\bar{t}^1)(t^2+\bar{t}^2)(u^3+\bar{u}^3))^{-1/2}.
\label{nonchiralK}
\end{equation}
The superpotential is given by
\begin{equation}
W = \hat{W}+ \frac{1}{2}\mu_{\alpha\beta}(M)C^{\alpha}C^{\beta} + \frac{1}{6}Y_(M){\alpha\beta\gamma}C^{\alpha\beta\gamma}+\cdots
\end{equation}
while the minimum of the F part of the tree-level supergravity
scalar potential $V$ is given by
\begin{equation}
V(M,\bar{M}) = e^G(G_M K^{MN} G_N -3) = (F^N K_{NM} F^M-3e^G),
\end{equation}
where
$G_M=\partial_M G$ and $K_{NM}=\partial_N \partial_M K$, $K^{MN}$
is inverse of $K_{NM}$, and the auxiliary fields $F^M$ are given
by
\begin{equation}
F^M=e^{G/2} K^{ML}G_L. \label{aux}
\end{equation}
Supersymmetry is broken when some of the F-terms of the hidden sector fields $M$
acquire VEVs. This then results in soft terms being generated in
the observable sector. For simplicity, it is assumed in this
analysis that the $D$-term does not contribute (see
\cite{Kawamura:1996ex}) to the SUSY breaking.  Then the goldstino
is included by the gravitino via the superHiggs effect. The
gravitino then obtains a mass
\begin{equation}
m_{3/2}=e^{G/2},
\end{equation}
which we will take to be $\approx1$~TeV in the following. The
normalized gaugino mass parameters, scalar mass-squared
parameters, and trilinear parameters respectively may be given in
terms of the K\a"ahler potential, the gauge kinetic function, and
the superpotential as
\begin{eqnarray}
M_P &=& \frac{1}{2\mbox{Re}f_P}(F^M\partial_M f_P), \\ \nonumber
m^2_{PQ} &=& (m^2_{3/2} + V_0) - \sum_{M,N}\bar{F}^{\bar{M}}F^N\partial_{\bar{M}}\partial_{N}log(\tilde{K}_{PQ}), \\ \nonumber
A_{PQR} &=& F^M\left[\hat{K}_M + \partial_M log(Y_{PQR}) - \partial_M log(\tilde{K}_{PQ}\tilde{K}_{QR}\tilde{K}_{RP})\right],
\label{softterms}
\end{eqnarray}
where $\hat{K}_M$ is the K\a"ahler metric appropriate for branes
which are parallel on at least one torus, i.e. involving
non-chiral matter.  

The above formulas for the soft terms depend on the Yukawa couplings, via the superpotential.  An important consideration is whether or not this should cause any modification to the low-energy spectrum.  However, this turns out not to be the case since the Yukawas in the soft term formulas are not the same as the physical Yukawas, which arise from world-sheet instantons and are proportional to $exp({-A})$, where $A$ is the world-sheet area of the triangles formed by a triplet of intersections at which
the Standard Model fields are localized.  As we shall see in a later section, the physical Yukawa couplings in Type IIA depend on the K\a"ahler moduli and the open-string moduli.  This ensures that the Yukawa couplings present in the soft terms do not depend on either the complex-structure moduli or dilaton (in the supergravity basis).  Thus, the
Yukawa couplings will not affect the low-energy spectrum in the case of $u$-moduli dominant and mixed $u$ and $s$ dominant supersymmetry breaking.

To determine the SUSY soft breaking
parameters, and therefore the spectra of the models, we introduce
the VEVs of the auxiliary fields Eq. (\ref{aux}) for the
dilaton, complex and K\"ahler moduli \cite{Brignole:1993dj}:
\begin{eqnarray}
&& F^s=2\sqrt{3}C m_{3/2} {\rm Re}(s) \Theta_s e^{-i\gamma_s},
\nonumber \\
&&F^{\{u,t\}^i} = 2\sqrt{3}C m_{3/2}( {\rm Re}  ({u}^i) \Theta_i^u
e^{-i\gamma^u_i}+  {\rm Re} ({t}^i) \Theta_i^t
e^{-i\gamma_i^t}).
\end{eqnarray}
The factors $\gamma_s$ and $\gamma_i$ are the CP violating phases of the moduli, while
the constant $C$
is given by
\begin{equation}
C^2 = 1+ \frac{V_0}{3 m^2_{3/2}}.
\end{equation}
The goldstino is included in the gravitino by $\Theta_S$ in $S$
field space, and $\Theta_i$ parameterize the goldstino direction
in $U^i$ space,  where $\sum (|\Theta_i^u|^2 + |\Theta_i^t|^2) + |\Theta_s|^2 =1$. The
goldstino angle $\Theta_s$ determines the degree to which SUSY
breaking is being dominated by the dilaton $s$ and/or complex
structure ($u^i$) and K\"ahler ($t^i$) moduli.  As suggested earlier, we
will not consider the case of $t$-moduli dominant supersymmetry breaking 
as in this case, the soft terms are not independent of the Yukawa couplings.

\section{Gauge Coupling Unification}

The MSSM predicts the unification of the three gauge couplings at
an energy $\sim2.4\times10^{16}$~GeV. In intersecting D-brane
models, the gauge groups arise from different stacks of branes,
and so they will not generally have the same volume in the
compactified space. Thus, the gauge couplings are not
automatically unified, in contrast to heterotic models. For branes
wrapping cycles not invariant under $\Omega R$, the holomorphic
gauge kinetic function for a D6 brane stack $P$ is given by
Eq.~(\ref{kingauagefun}). where $u^i$ and $s$ are the complex
structure moduli and dilaton in the supergravity basis.

The gauge coupling constant associated with a stack P is given by
\begin{eqnarray}
g_{D6_P}^{-2} &=& |\mathrm{Re}\,(f_P)|.\label{idb:eq:gkf}
\end{eqnarray}
\noindent Thus, for the model under study the $SU(3)$ holomorphic
gauge function is identified with stack $a1$ and the $SU(2)$
holomorphic gauge function with stack $b$. The $Q_Y$ holomorphic
gauge function is then given by taking a linear combination of the
holomorphic gauge functions from all the stacks. Note
that we have absorbed a factor of $1/2$ in the definition of $Q_Y$
so that the electric charge is given by $Q_{em} = T_3 + Q_Y$. In
this way, it is found~\cite{Blumenhagen:2003jy} that
\begin{equation}
f_Y = \frac{1}{6}f_{a1} + \frac{1}{2}f_{a2} + \frac{1}{2}f_{c1} + \frac{1}{2}f_{c2}.
\end{equation}
Recalling that the complex structure moduli $U^i$ are obtained from
the supersymmetry conditions, we have
for the present model
\begin{equation}
U^1 = 3i~,~~ U^2 = i~,~~ U^3 = -1 + i~.~\,
\end{equation}

Thus, we find that the tree-level MSSM gauge couplings will
 be automatically unified at the string scale
\begin{equation}
g^2_{s} = g^2_{w} = \frac{5}{3}g^2_Y = \left[\frac{e^{-\phi_4}}{2\pi} \frac{\sqrt{6}}{4}\right]^{-1}.
\end{equation}
Even though the gauge couplings are unified, this does not fix the
actual value of the couplings as these still depend upon the value
taken by the four-dimensional dilaton $\phi_4$. In order for the
gauge couplings to have the value observed for the MSSM ($g^2_{\rm unification}
\approx 0.511$), we must choose $\phi_4 = -3$ such that
$e^{-\phi_4} \approx 20$, which fixes the string scale as
\begin{equation}
M_{St} = \pi^{1/2} e^{\phi_4} M_{Pl} \approx 2.1 \times 10^{17}~\mbox{GeV},
\end{equation}
where $M_{Pl}$ is the reduced Planck scale.

It should be kept in mind that values
given for the gauge couplings at the string scale are only the
{\it tree-level} results. There are one-loop threshold
corrections arising from the $N=1$ and $N=2$ open string
sectors~\cite{Lust:2003ky} which may alter these results. 
In addition, there is exotic matter charged under both
observable and hidden sector gauge groups, which
are expected to pick up large masses, but could
still affect the running of the gauge couplings.  

\section{Confinement of the Hidden Sector Fields}  
\begin{figure}[h]
    \centering
        \includegraphics[width=0.45\textwidth]{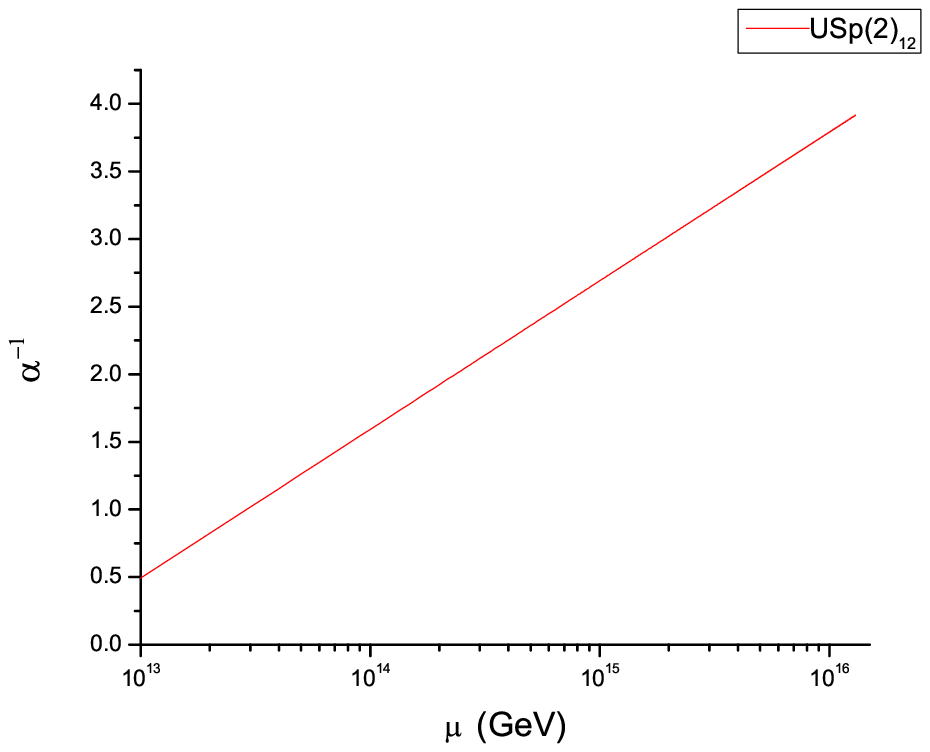}
        \includegraphics[width=0.45\textwidth]{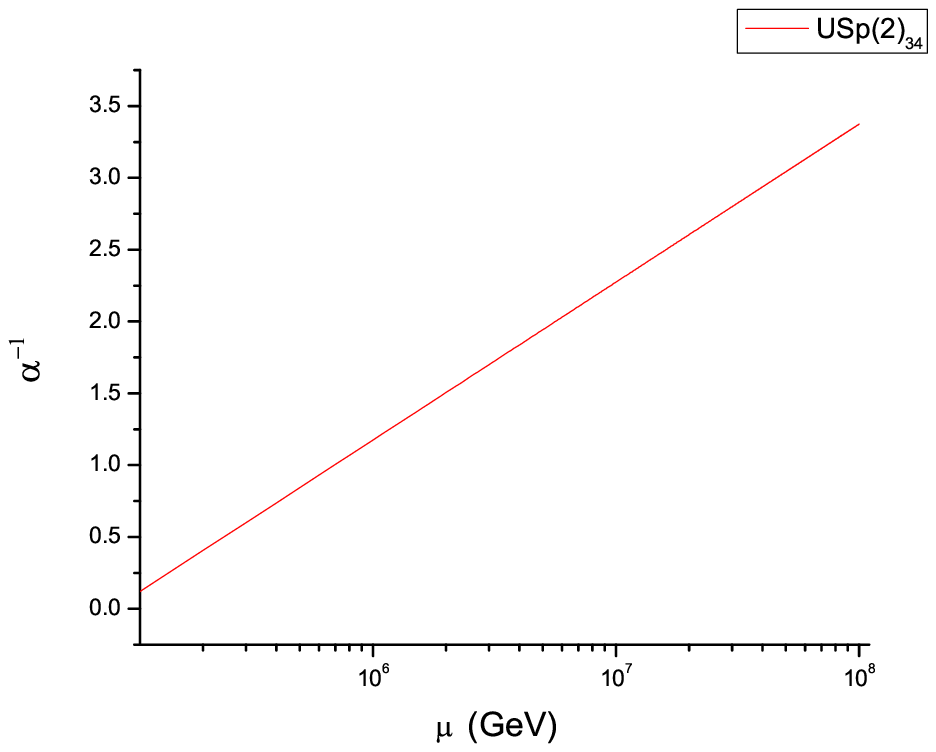}
    \caption{RGE running of the gauge coupling for $USp(2)_1/USp(2)_2$, $USp(2)_3/USp(2)_4$
    hidden sector gauge groups, which become confining at $\approx3\cdot10^{13}$~GeV and $\approx6\cdot10^{5}$ respectively.}
    \label{fig:USp}
\end{figure}

In addition to the matter content of the MSSM, there is also matter
charged under the hidden sector $USp(2)$ gauge groups. These
states will generally have fractional electric charges, similar
to the so-called \lq
cryptons\rq~\cite{Ellis:1990iu,Benakli:1998ut,Ellis:2004cj,Ellis:2005jc}.
Obviously, no such matter is observed in the low-energy spectrum
so these exotic states must receive a large mass.  Such a mass may
arise if the hidden sector gauge couplings are asymptotically free
and become confining at some high energy. Indeed, in the present
case we find that the $\beta$-functions for the $USp(2)$ groups
are all negative~\cite{CLL}, 
\begin{equation}
 \beta_{USp(2)_1}=\beta_{USp(2)_2}=\beta_{USp(2)_3}=\beta_{USp(2)_4}= -3,
\end{equation}
where we consider all of the chiral
exotic particles present even though it is expected that these
states will decouple as discussed previously.
From the holomorphic gauge kinetic function, the gauge
couplings are found to take the values
\begin{eqnarray}
g^2_{USp(2)_1} = g^2_{USp(2)_2} \approx 3, \\ \nonumber
g^2_{USp(2)_3} = g^2_{USp(2)_4} \approx 1.
\end{eqnarray}
at the string scale.  
We may then straightforwardly run these couplings to low-energy
energy via the one-loop RGE equations,
\begin{equation}
\frac{1}{g^2(\mu)} = \frac{1}{g^2_{M_{st}}} - \frac{1}{8\pi^2}\beta~\mbox{ln}\left(\frac{\mu}{M_{st}}\right),
\end{equation}
where we find that the couplings  for the $USp(2)_1$ and $USp(2)_2$
hidden sector groups will become strong at a scale
$\sim3\cdot10^{13}$~GeV, while the couplings for the $USp(2)_3$ and
$USp(2)_4$ groups will become strong around $\sim7\cdot10^{5}$~GeV
as shown in Figure~\ref{fig:USp}.

We should note that it is also possible to decouple the chiral exotic states in the manner  discussed in section II.

\section{Soft Terms and Superpartner Spectra}

Next, we turn to our attention to the soft supersymmetry breaking
terms at the GUT scale defined in Eq.~(\ref{softterms}).  In the
present analysis, not all the F-terms of the moduli get VEVs for simplicity, as
in \cite{Font:2004cx, Kane:2004hm}. As discussed earlier, we will assume that $F^t_i=0$ so that the soft terms have
no dependence on the physical Yukawa couplings. Thus, we consider two cases:
\begin{enumerate}
    \item The $u$-moduli dominated SUSY breaking where both the cosmological
    constant $V_0$ and the goldstino angle are set to zero, such that $F^s=F^{t^i}=0$.
    \item The $u$ and $s$-moduli SUSY breaking where the cosmological constant
  $V_0 = 0$ and $F^s \neq 0$.
\end{enumerate}

\subsection{SUSY breaking with $u$-moduli dominance}

For this case we take $\Theta_s = 0$ so that the $F$-terms are parameterized by the expression
\begin{equation}
F^{u^i} = \sqrt{3}m_{3/2}(u^i + \bar{u}^i)\Theta_i e^{-i\gamma_i},
\label{auxfields}
\end{equation}
where $i = 1\mbox{,} 2\mbox{,} 3$ and with $\sum |\Theta_i|^2 =
1$. With this parametrization, the gaugino mass terms for a stack
$P$ may be written as
\begin{eqnarray}
M_P=\frac{-\sqrt{3}m_{3/2}}{\mbox{Re}
f_P}\sum_{j=1}^3 \left(\mbox{Re} u^j\,\Theta_j\,
e^{-i\gamma_j}\,n^j_Pm^k_Pm^l_P\right) \qquad
(j,k,l)=(\overline{1,2,3}). \label{eq:idb:gaugino}
\end{eqnarray}
The Bino mass parameter is a linear combination of the
gaugino mass for each stack,
\begin{equation}
M_Y = \frac{1}{f_Y}\sum_P c_P M_P
\end{equation}
where the the coefficients $c_P$ correspond to the linear
combination of $U(1)$ factors
which define the hypercharge, $U(1)_Y = \sum c_P U(1)_P$.

For the trilinear parameters, we have
\begin{eqnarray}
A_{PQR}&=&-\sqrt{3}m_{3/2}\sum_{j=1}^3 \left[
\Theta_je^{-i\gamma_j}\left(1+(\sum_{k=1}^3
 \xi_{PQ}^{k,j}\Psi(\theta^k_{PQ})-\frac{1}{4})+(\sum_{k=1}^3
 \xi_{RP}^{k,j}\Psi(\theta^k_{RP})-\frac{1}{4})
\right)\right]  \nonumber \\
&&+\frac{\sqrt{3}}{2}m_{3/2}{\Theta}_{3}e^{-i{\gamma}_1}
\end{eqnarray}
where $P$,$Q$, and $R$ label the stacks of branes whose mutual
intersections define the fields present in the corresponding
trilinear coupling and the angle differences are defined as
\begin{equation}
\theta_{PQ} = \theta_Q - \theta_P.
\end{equation}
We must be careful when dealing with cases where the angle difference is
negative.  Note for the present model, there is always either one
or two of the $\theta_{PQ}$ which are negative.  Let us define the
parameter
\begin{equation}
\eta_{PQ} = \mbox{sgn}(\prod_i \theta_{PQ}^i),
\end{equation}
such that $\eta_{PQ} = -1$ indicates that only one of the angle
differences are negative while $\eta_{PQ} = +1$ indicates that two
of the angle differences are negative.

Finally, the squark and slepton (1/4 BPS) scalar mass-squared
parameters are given by
\begin{eqnarray}
m^2_{PQ}= m_{3/2}^2\left[1-3\sum_{m,n=1}^3
\Theta_m\Theta_ne^{-i(\gamma_m-\gamma_n)}\left(
\frac{{\delta}_{mn}}{4}+ \sum_{j=1}^3 (\xi^{j,m\bar
n}_{PQ}\Psi(\theta^j_{PQ})+
 \xi^{j,m}_{PQ}\xi^{j,\bar n}_{PQ}\Psi'(\theta^j_{PQ}))\right)
\right].
\end{eqnarray}

The functions $\Psi(\theta_{PQ})=\frac{\partial \ln
(e^{-\phi_4}\tilde{K}_{PQ})}{\partial \theta_{PQ}}$ in the above
formulas defined for $\eta_{PQ}=-1$ are
\begin{eqnarray}
\mbox{if} \ \theta_{PQ} < 0&:& \\ \nonumber
\Psi(\theta^j_{PQ})&=&
-\gamma_E+\frac{1}{2}\frac{d}{d{\theta}^j_{PQ}}\,\ln{\Gamma(-\theta^j_{PQ})}-
\frac{1}{2}\frac{d}{d{\theta}^j_{PQ}}\,\ln{\Gamma(1+\theta^j_{PQ})}+\ln(t^j+\bar t^j)\\ \nonumber
\mbox{if} \ \theta_{PQ} > 0&:& \\ \nonumber
\Psi(\theta^j_{PQ})&=&
-\gamma_E+\frac{1}{2}\frac{d}{d{\theta}^j_{PQ}}\,\ln{\Gamma(1-\theta^j_{PQ})}-
\frac{1}{2}\frac{d}{d{\theta}^j_{PQ}}\,\ln{\Gamma(\theta^j_{PQ})}+\ln(t^j+\bar t^j),
\label{eqn:Psi1}
\end{eqnarray}
and for $\eta_{PQ}=+1$ are
\begin{eqnarray}
\mbox{if} \ \theta_{PQ} < 0&:& \\ \nonumber
\Psi(\theta^j_{PQ})&=&
\gamma_E+\frac{1}{2}\frac{d}{d{\theta}^j_{PQ}}\,\ln{\Gamma(1+\theta^j_{PQ})}-
\frac{1}{2}\frac{d}{d{\theta}^j_{PQ}}\,\ln{\Gamma(-\theta^j_{PQ})}-\ln(t^j+\bar t^j)\\ \nonumber
\mbox{if} \ \theta_{PQ} > 0&:& \\ \nonumber
\Psi(\theta^j_{PQ})&=&
\gamma_E+\frac{1}{2}\frac{d}{d{\theta}^j_{PQ}}\,\ln{\Gamma(\theta^j_{PQ})}-
\frac{1}{2}\frac{d}{d{\theta}^j_{PQ}}\,\ln{\Gamma(1-\theta^j_{PQ})}-\ln(t^j+\bar t^j).
\label{eqn:Psi2}
\end{eqnarray}
The function $\Psi'(\theta_{PQ})$ is just the derivative
\begin{eqnarray}
\Psi'(\theta^j_{PQ})&=&
\frac{d\Psi(\theta^j_{PQ})}{d \theta^j_{PQ}},
\label{eqn:Psip}
\end{eqnarray}
and ${\theta}^{j,k}_{PQ}$ and ${\theta}^{j,k\bar{l}}_{PQ}$ are
defined~\cite{Kane:2004hm} as

\begin{equation}
{\xi}^{j,k}_{PQ} \equiv (u^k+\bar u^k)\,\frac{\partial
\theta^j_{PQ}}{\partial u^k}= \left\{\begin{array}{cc}
 \left[-\frac{1}{4\pi}
 \sin(2\pi\theta^j)
 \right]^P_Q & \mbox{ when }j=k  \vspace*{0.6cm} \\
 \left[\frac{1}{4\pi}
\sin(2\pi\theta^j)
 \right]^P_Q & \mbox{ when }j\neq k,
\end{array}\right.\label{idb:eq:dthdu}
\end{equation}

\begin{equation}
{\xi}^{j,k\bar{l}}_{PQ} \equiv (u^k+\bar
u^k)(u^l+\bar u^l)\,\frac{\partial^2 \theta^j_{PQ}}{\partial
u^k\partial\bar u^l}= \left\{\begin{array}{cc} \frac{1}{16\pi}
  \left[ \sin(4\pi\theta^j)+4\sin(2\pi\theta^j)
 \right]^P_Q &
   \mbox{when }j=k=l  \vspace*{0.6cm} \\
 \frac{1}{16\pi}  \left[
 \sin(4\pi\theta^j)-4\sin(2\pi\theta^j)
 \right]^P_Q &
   \mbox{when }j\neq k=l  \vspace*{0.6cm} \\
 -\frac{1}{16\pi}\left[
 \sin(4\pi\theta^j)
 \right]^P_Q &
   \mbox{ when }j=k\neq l\mbox{ or } j=l\neq k \vspace*{0.4cm} \\
 \frac{1}{16\pi}\left[
\sin(4\pi\theta^j)
 \right]^P_Q &
   \mbox{when }j\neq k\neq l\neq j.
\end{array}\right.\label{idb:eq:dth2du}
\end{equation}

Note that the only explicit dependence of the soft terms on the
$u$ and $s$ moduli is in the gaugino mass parameters.  The
trilinears and scalar mass-squared values depend explicitly only
on the angles.  However, there is an implicit dependence on the
complex structure moduli via the angles made by each D-brane with respect to the
orientifold planes.

In contrast to heterotic string models, the gaugino and scalar
masses are typically not universal in intersecting D-brane
constructions, although in the present case, there is some partial
universality of the scalar masses due to the Pati-Salam
unification at the string scale. In particular, the trilinear $A$
couplings are found to be equal to a universal parameter, $A_0$
and the left-handed and right-handed squarks and sleptons
respectively are degenerate. The Higgs states arise from the
non-chiral sector due to the fact that stacks $b$, $c1$, and $c2$
are parallel on the third torus.  The appropriate K\a"ahler metric
for these states is given by Eq.~(\ref{nonchiralK}).  Thus, the
Higgs scalar mass-squared values are found to be
\begin{equation}
m^2_H = m^2_{3/2}\left(1-\frac{3}{2}\left|\Theta_3\right|^2\right).
\end{equation}

The complex structure moduli $u^i$, and the four-dimensional
dilaton $\phi_4$ are fixed by the supersymmetry conditions and
gauge coupling unification respectively. The K\a"ahler modulus on
the first torus $t^1$ will be chosen to be consistent with the Yukawa mass
matrices calculated in the next section. Thus, the free parameters which remain are
$\Theta_{1,2}$, $\mbox{sgn}(\Theta_3)$, $t^2$, $t^3$, the phases
$\gamma_i$, and the gravitino mass $m_{3/2}$. In order to
eliminate potential problems with electric dipole moments (EDM's) for the 
neutron and electron,
 we set $\gamma_i=0$.  In
addition, we set the K\a"ahler moduli on the second and third tori
equal to one another, $\mbox{Re}(t^2)=\mbox{Re}(t^3)=0.5$ and take
the gravitino mass $m_{3/2} \sim 1$~TeV.  Note that the soft terms
only have a weak logarithmic dependence on the K\a"ahler moduli.

We constrain the parameter space such that
neither the Higgs nor the squark and slepton scalar masses are
tachyonic at the high scale, as well as imposing the unitary condition
$\Theta_1^2+\Theta_2^2+\Theta_3^3=1$.  In particular, we require
$\Theta_3^2 \leq 2/3$, or equivalently $\Theta_1^2+\Theta_2^2 \geq
1/3$.

To determine the soft terms and superpartner spectra at the low
energy scale, we run the RGE's
down from the high scale using the code {\tt
SuSpect}~\cite{Djouadi:2002ze}. In principle, we should be able to
determine tan~$\beta$, and the $\mu$ and $B$ parameters directly from the model. For 
the present construction, there is in fact a $\mu$ parameter, whose real part
corresponds geometrically 
to the separation between stacks $b$ and $c$.  In the absence of any effects
which stabilize the open-string moduli, there are corresponding flat directions in
the moduli space.  Indeed, the calculation of the Yukawa couplings in the next section
will exploit this freedom.  Thus, the effective $\mu$-term cannot be calculated until
the moduli stablization issue has been addressed.  Essentially the same considerations
apply for a determination of tan~$\beta$, which depends upon the explicit values for the
neutral component VEV of the Higgs fields, up to an over-all constant.  The overall issue
of moduli stablization is discussed in a later section.  

For the present work, we will fix these values via the
requirement for electroweak symmetry breaking (EWSB), in a similar fashion to~\cite{Kane:2004hm}. We also
choose $\mu > 0$ as favored by ($g-2$) and take tan~$\beta$ as a
free parameter. We use the value for the top quark mass $m_t = 170.9$~GeV.  Then, knowing the low energy spectra, we can then
determine the corresponding neutralino relic density.  To
calculate this, we use the code {\tt
MicrOMEGAS~\cite{Belanger:2004yn}}.  Some points in the parameter
space which give the observed dark matter density are shown in
Table~\ref{intnumPS} where we have fixed $\Theta_s = 0$.  

\begin{table}[h]
\footnotesize
\caption{Supersymmetry breaking soft terms at the string scale and resulting neutralino relic density for
some specific choices of goldstino angles, with $\Theta_s = 0$.}
\begin{center}
\begin{tabular}{|c|c|c|c|c|c|c|c|c|c|c|c|c|c|c|c|c|c|c|} \hline
                 $\Theta_1$&$\Theta_2$&$\Theta_3$&$M_{\tilde{G}}$~(GeV)&$M_{\tilde{W}}$~(GeV)&$M_{\tilde{B}}$~(GeV)&$m_H$~(GeV)&$m_L$~(GeV)&$m_R$~(GeV)&$A_t$~(GeV)& LSP & $\Omega h^2$ \\ \hline\hline
$-0.610$ & $0.290$ & $0.737$    & $ 889$ & $251$ & $422$ & $429$ & $ 963$  & $466$ &    $ 676$ & $\tilde{B}$ & $0.115$ \\ $-0.610$ & $0.380$  & $0.695$ & $ 931$ & $329$ & $416$ & $524$ & $ 989$  & $478$ &  $ 610$ & $\tilde{B}$ & $0.108$ \\
$-0.600$ & $0.470$  & $0.647$ & $ 967$ & $407$ & $411$ & $609$ & $1005$  & $492$ &  $ 531$ & $\tilde{B}$ & $0.113$ \\
$-0.100$ & $0.870$  & $0.482$ & $1171$ & $753$ & $667$ & $806$ & $ 684$  & $521$ &  $-312$ & $\tilde{B}$ & $0.118$ \\
$0.0400$ & $0.600$  & $0.799$ & $1211$ & $519$ & $920$ & $205$ & $ 510$  & $590$ &  $-222$ & $\tilde{B}$ & $0.105$ \\
$0.110$ & $-0.570$ & $0.814$ & $    211$ & $-493$ & $564$ & $74$ & $ 561$  & $631$ &  $ 431$ & $\tilde{B}/\tilde{H}$ &
$0.107$ \\
$0.150$  & $0.660$ & $0.736$ &  $1209$ & $571$ & $944$ & $432$ & $ 529$  & $512$ &  $-397$ & $\tilde{B}$ & $0.108$ \\
$0.010$  & $0.810$ & $0.586$ &  $1209$ & $701$ & $793$ & $695$ & $ 603$  & $510$ &  $-371$ & $\tilde{B}$ & $0.094$ \\
\hline\hline

\end{tabular}
\end{center}
\label{intnumPS}
\end{table}

\begin{table}[htb]

\footnotesize

\renewcommand{\arraystretch}{1.0}

\caption{Low energy supersymmetric particles and their masses (in GeV) for $\Theta_1 = -0.610$,
$\Theta_2 = 0.290$, $\Theta_3 = 0.737$, and $\Theta_s = 0.$ with tan~$\beta=46$.}

\label{SSpectrum1}

\begin{center}

\begin{tabular}{|c|c|c|c|c|c|c|c|c|c|}\hline

$h^0$ & $H^0$ & $A^0$ & $H^{\pm}$ & ${\tilde g}$ & $\chi_1^{\pm}$
& $\chi_2^{\pm}$ & $\chi_1^{0}$ & $\chi_2^{0}$  \\ \hline

116.89  & 826.80   & 826.81    & 831.05   & 1985  & 192.3    & 1115.   & 173.3
&  192.3      \\ \hline

$\chi_3^{0}$ & $\chi_4^{0}$ & ${\tilde t}_1$ & ${\tilde t}_2$ &
${\tilde u}_1/{\tilde c}_1$ & ${\tilde u}_2/{\tilde c}_2$ &
${\tilde b}_1$ & ${\tilde b}_2$ & \\ \hline

-1113.  & 1114  & 1477  & 1789 & 1949 & 1761 & 1638 & 1791 &\\ \hline

${\tilde d}_1/{\tilde s}_1$ & ${\tilde d}_2/{\tilde s}_2$ &
${\tilde \tau}_1$ & ${\tilde \tau}_2$ & ${\tilde \nu}_{\tau}$ &
${\tilde e}_1/{\tilde \mu}_1$ & ${\tilde e}_2/{\tilde \mu}_2$ &
${\tilde \nu}_e/{\tilde \nu}_{\mu}$ & \\ \hline

1950 & 1760 & 189.0 & 928.2 & 919.5 & 973.3 & 488.0 & 970.1 & \\

\hline

\end{tabular}

\end{center}

\end{table}

\begin{table}[htb]

\footnotesize

\renewcommand{\arraystretch}{1.0}

\caption{Low energy supersymmetric particles and their masses (in GeV) for $\Theta_1 = 0.04$,
$\Theta_2 = 0.60$, $\Theta_3 = 0.799$, and $\Theta_s = 0$ with tan~$\beta=46$.}

\label{SSpectrum2}

\begin{center}

\begin{tabular}{|c|c|c|c|c|c|c|c|c|c|}\hline

$h^0$ & $H^0$ & $A^0$ & $H^{\pm}$ & ${\tilde g}$ & $\chi_1^{\pm}$
& $\chi_2^{\pm}$ & $\chi_1^{0}$ & $\chi_2^{0}$  \\ \hline

118.82  & 1057.1  & 1057.1    & 1060.4  & 2616.  & 414.6    & 1478.   & 390.4
&  414.6      \\ \hline

$\chi_3^{0}$ & $\chi_4^{0}$ & ${\tilde t}_1$ & ${\tilde t}_2$ &
${\tilde u}_1/{\tilde c}_1$ & ${\tilde u}_2/{\tilde c}_2$ &
${\tilde b}_1$ & ${\tilde b}_2$ & \\ \hline

-1475.  & 1477  & 1955  & 2114 & 2318 & 2335 & 2071 & 2184 &\\ \hline

${\tilde d}_1/{\tilde s}_1$ & ${\tilde d}_2/{\tilde s}_2$ &
${\tilde \tau}_1$ & ${\tilde \tau}_2$ & ${\tilde \nu}_{\tau}$ &
${\tilde e}_1/{\tilde \mu}_1$ & ${\tilde e}_2/{\tilde \mu}_2$ &
${\tilde \nu}_e/{\tilde \nu}_{\mu}$ & \\ \hline

2319 & 2330 & 474.4 & 689.6 & 574.9 & 620.1 & 679.0 & 615.1 & \\

\hline

\end{tabular}

\end{center}

\end{table}

\begin{figure}
    \centering
        \includegraphics[width=0.75\textwidth]{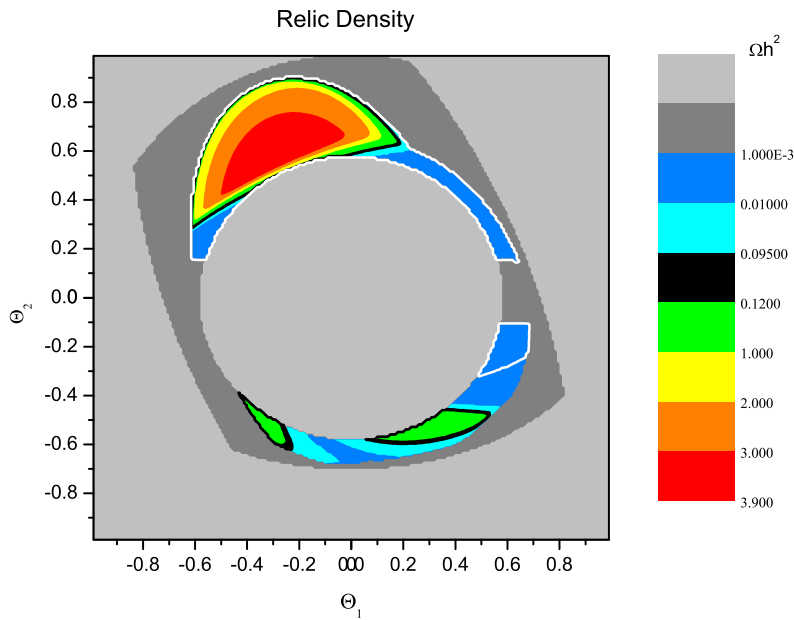}
    \caption{Contour map of the dark matter density for tan~$\beta=46$, $\theta_3 > 0$, $\Theta_s = 0$ and $\mu > 0$.  The areas in black denote regions
    where $0.09 \leq \Omega h^2 \leq 0.12$ with a gravitino mass $m_{3/2} = 1$~TeV. The light gray regions are excluded because they do not satisfy the constraints on the soft terms at high scale.  The dark grey regions denote regions where either the neutralino is not the LSP, mass limits are not satisfied or for which there is no RGE solution.  Regions inside the white contour satisfy the LEP limit, $m_h > 114$~GeV.}
    \label{fig:DMtb45}
\end{figure}

A contour plot of the dark matter density is displayed in
Fig~\ref{fig:DMtb45} with $m_{3/2}=1$~TeV, tan~$\beta=46$ and for
$\Theta_3>0$. It can be seen that only small regions of the
allowed parameter space can produce the observed dark matter
density, which are indicated on the plot as dark bands. 
Regions which 
do not satisfy the constraints
\begin{equation}
\Theta_1^2+\Theta_2^2+\Theta_3^3=1
\end{equation}
and
\begin{equation}
\sqrt{\Theta_1^2+\Theta_2^2 }\geq
\frac{1}{\sqrt{3}}
\end{equation}
are indicated on
the plots as the light gray shaded areas.
The dark gray areas indicate regions of
the parameter space where either the neutralino
is not the LSP, LEP superpartner mass limits are not
satisfied, or for which there is no RGE solution.
The viable
parameter space may be further constrained by imposing the LEP
limit on the lightest CP-even Higgs mass, $m_h \geq 114$~GeV. Regions satisfying
this bound are contained within the white contour on the plot.
Essentially, this rules out regions of the parameter space with a
mixed Bino/Higgsino LSP as the Higgs mass for these regions of the
parameter space is always below $114$~GeV. A sampling of some of
the SUSY spectra for some of the cases of Table~\ref{intnumPS}
are shown in Tables~\ref{SSpectrum1} and~\ref{SSpectrum2}.

\begin{figure}[ht]
    \centering
        \includegraphics[width=0.47\textwidth]{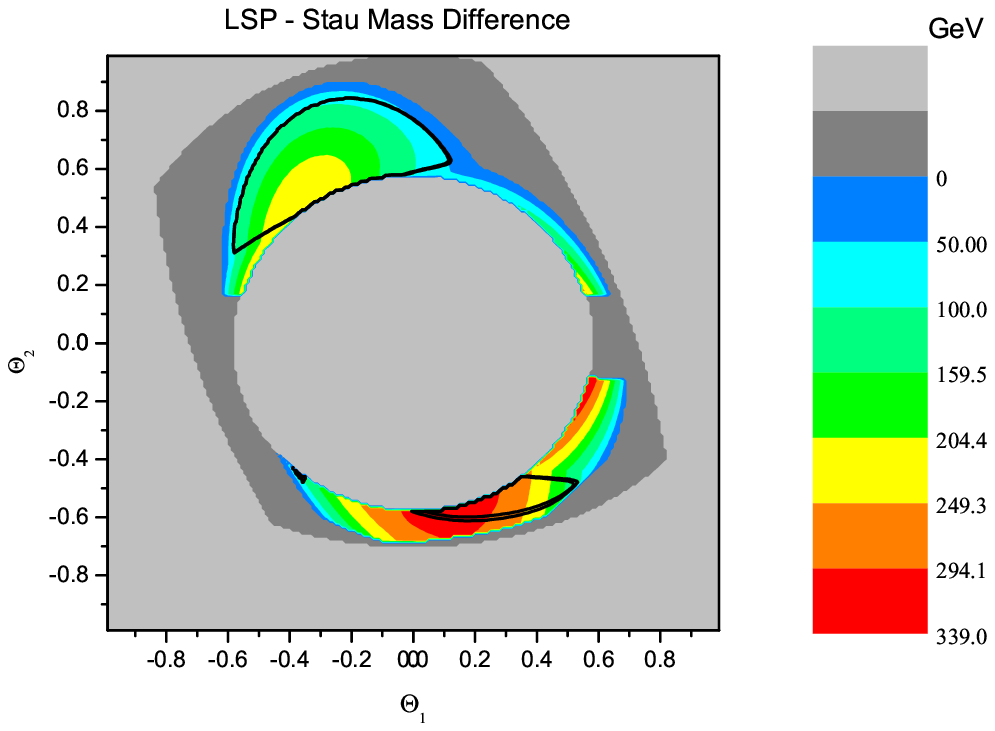}
        \includegraphics[width=0.47\textwidth]{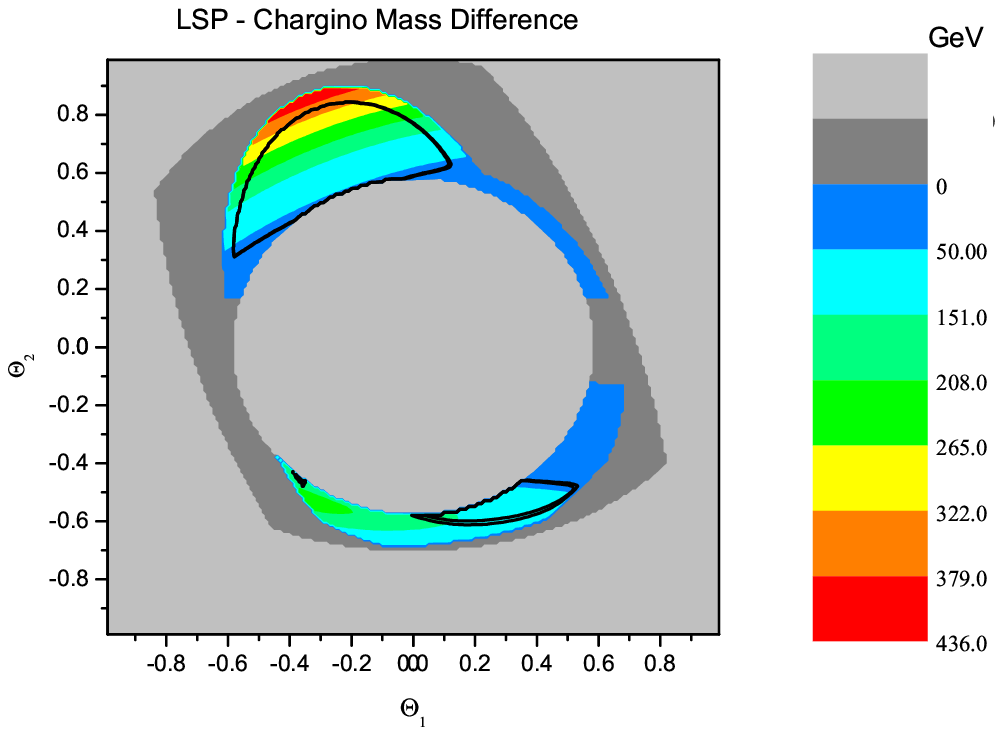}
    \caption{Contour map of the neutralino-stau mass difference and neutralino-chargino mass difference for tan~$\beta=46$, $\theta_3 > 0$, $\Theta_s = 0$ and $\mu > 0$.  The areas in black denote regions
    where $0.095 \leq \Omega h^2 \leq 0.125$ with a gravitino mass $m_{3/2} = 1$~TeV. The light gray regions are excluded because they do not satisfy the constraints on the soft terms at high scale.  The dark grey regions denote regions where either the neutralino is not the LSP, mass limits are not satisfied or for which there is no RGE solution.  The dark matter density within the allowed range coincides with the $\tilde{\chi}^0_1~\tilde{\tau}$ and $\tilde{\chi}^0_1~\tilde{\chi}^{\pm}/\tilde{\chi}^0_1~\tilde{\chi}^{0}_2$ coanhiliation regions where the stau and/or chargino/next-to-lightest neutralino mass is slightly bigger
    than the lightest neutralino mass.}
    \label{fig:Massdiff}
\end{figure}

It can be seen from Fig.~{\ref{fig:Massdiff} that regions of the corresponding
parameter space correspond to cases where the lightest
neutralino mass is very close to the either or both the lightest
chargino/next-to-lightest neutralino $\tilde{\chi}^{\pm}_1/\tilde{\chi}^{0}_2$ (which are mass degenerate)
and light stau $\tilde{\tau}_1$ mass.  In other words, the observed dark matter
density is obtained close to the $\tilde{\chi}^0_1~\tilde{\tau}_1$
and/or $\tilde{\chi}^0_1~\tilde{\chi}^{\pm}_1/\tilde{\chi}^0_1~\tilde{\chi}^{0}_2$
coanhilation regimes.

The lower bound on the gravitino mass may be estimated for fixed
tan~$\beta$ by lowering $m_{3/2}$ until there are no regions
of the parameter space for which the Higgs mass satisfies the LEP
limit.  In this way, it is found that the lower bound satisfies
$425~\mbox{GeV} \leq m_{3/2} \leq 450$~GeV for tan~$\beta=46$.
This can be seen in Figs.~\ref{fig:DMHiggs475GeV}, where the
region of the parameter space with a Higgs mass above $114$~GeV
begins to shrink dramatically.  For $m_{3/2}= 450$~GeV, only very
small regions of the viable parameter space results in a Higgs mass
above $114$~GeV.  For $m_{3/2}=425$~GeV, there is no region of the
parameter space above this limit.  
Similar results hold for other values of tan $\beta$.

\begin{figure}[ht]
    \centering
        \includegraphics[width=0.47\textwidth]{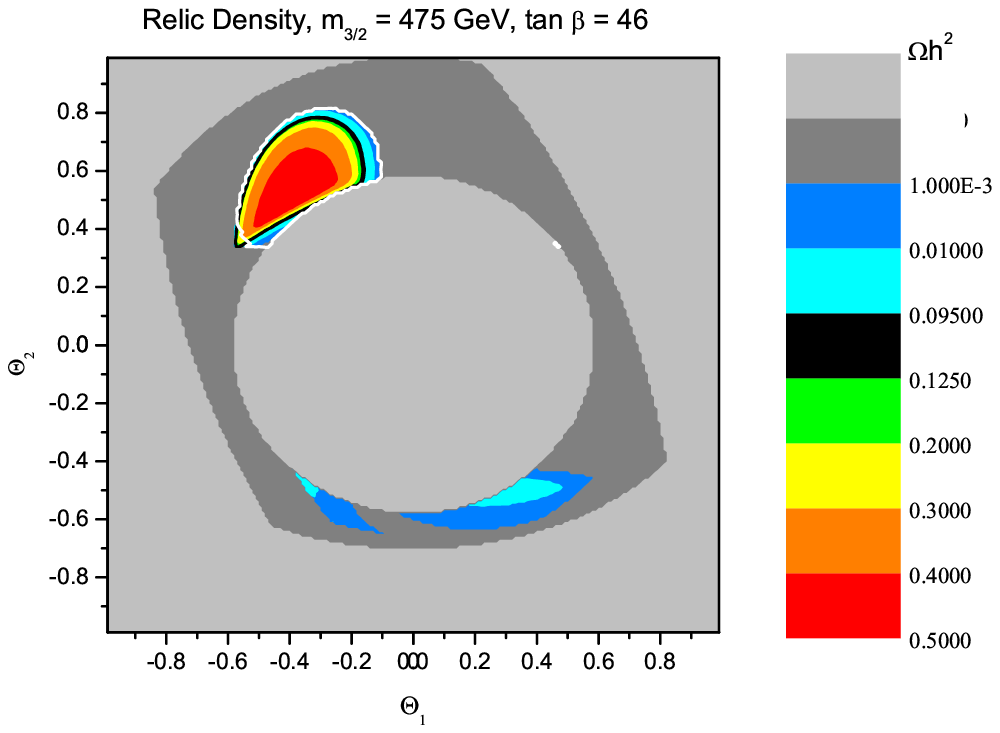}
         \includegraphics[width=0.47\textwidth]{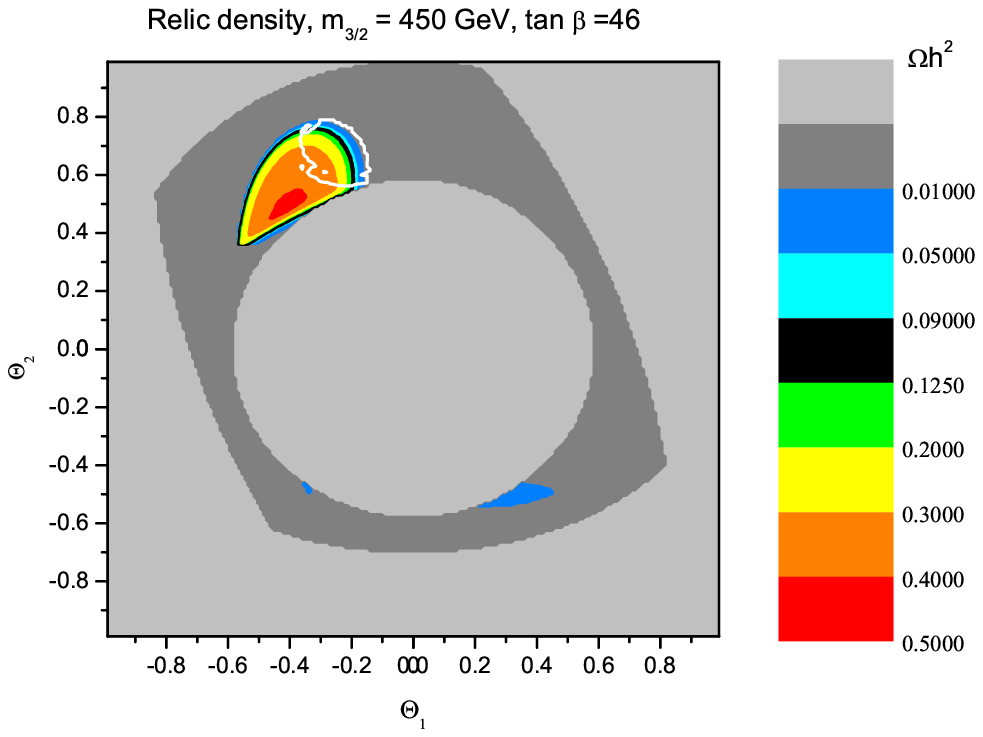}
    \caption{Contour maps of the relic density for tan~$\beta=46$, $\theta_3 > 0$, $\Theta_s=0$ and $\mu > 0$ with a gravitino mass $m_{3/2} = 475$~GeV and $m_{3/2} = 450$~GeV respectively.  The areas in black denote regions
    where $0.09 \leq \Omega h^2 \leq 0.12$. The light gray regions are excluded due to tachyonic scalar masses at the high scale.  The dark grey regions denote regions where LEP superpartner mass limits are not satisfied or there is no RGE solution.
    Regions inside the white contour satisfy the LEP limit, $m_h > 114$~GeV.  The viable region of the parameter
    space is much smaller for $m_{3/2}=450$~GeV and dissappears for $m_{3/2}=425$~GeV, which constrains
    the minimum gravitino mass to be in the range $425~\mbox{GeV}< m_{3/2} \leq 450~\mbox{GeV}$. }
    \label{fig:DMHiggs475GeV}
\end{figure}

\subsection{SUSY breaking via $u$-moduli and dilaton $s$}

Next, we allow the dilaton $s$ to obtain a non-zero VEV as well as the $u$-moduli.
To do this, we parameterize the $F$-terms as
\begin{equation}
F^{u^i,s} = \sqrt{3}m_{3/2}[(s + \bar{s})\Theta_s e^{-i\gamma_s} + (u^i + \bar{u}^i)\Theta_i e^{-i\gamma_i}]
\end{equation}

Clearly, this is a more complicated situation with a much larger set
of values over which to scan.  The formula for
the gaugino mass associated with each stack now becomes
\begin{eqnarray}
M_P=\frac{-\sqrt{3}m_{3/2}}{\mbox{Re} f_P}\left[\left(\sum_{j=1}^3
\mbox{Re} (u^j)\,\Theta_j\,
e^{-i\gamma_j}\,n^j_Pm^k_Pm^l_P2^{-(\beta_k +\beta_l)}\right) +
\Theta_s\mbox{Re}(s) e^{-i\gamma_0}n_P^1\,n_P^2\,n_P^3\, \right],
\\ \nonumber \qquad (j,k,l)=(\overline{1,2,3}).
\end{eqnarray}
As before, the Bino mass parameter is a linear combination of the
gaugino mass for each stack, and the coefficients corresponding to
the linear combination of $U(1)$ factors define the hypercharge.

The trilinear parameters generalize as
\begin{eqnarray}
A_{PQR}&=&-\sqrt{3}m_{3/2}\sum_{j=0}^3 \left[
\Theta_je^{-i\gamma_j}\left(1+(\sum_{k=1}^3
 \xi_{PQ}^{k,j}\Psi(\theta^k_{PQ})-\frac{1}{4})+(\sum_{k=1}^3
 \xi_{RP}^{k,j}\Psi(\theta^k_{RP})-\frac{1}{4})
\right)\right] \nonumber \\
&&+\frac{\sqrt{3}}{2}m_{3/2}({\Theta}_{3}e^{-i{\gamma}_1} +
\Theta_s e^{-i{\gamma}_s}),
\end{eqnarray}
where $\Theta_0$ corresponds to $\Theta_s$  and there is a
contribution from the dilaton via the Higgs (1/2 BPS) 
K\a"ahler metric, which also gives an additional contribution to
the Higgs scalar mass-squared values:
\begin{equation}
m^2_H = m^2_{3/2}\left[1-\frac{3}{2}(\left|\Theta_3\right|^2+\left|\Theta_s\right|^2)\right].
\end{equation}

Finally, the squark and slepton (1/4 BPS) scalar mass-squared
parameters are given as before by
\begin{eqnarray}
m^2_{PQ}= m_{3/2}^2\left[1-3\sum_{m,n=0}^3
\Theta_m\Theta_ne^{-i(\gamma_m-\gamma_n)}\left(
\frac{{\delta}_{mn}}{4}+ \sum_{j=1}^3 (\xi^{j,m\bar
n}_{PQ}\Psi(\theta^j_{PQ})+
 \xi^{j,m}_{PQ}\xi^{j,\bar n}_{PQ}\Psi'(\theta^j_{PQ}))\right)
\right],
\end{eqnarray}
where we now also include the $\Theta_s = \Theta_0$ in the sum. The
functions $\Psi(\theta_{PQ})$ and $\Psi'(\theta_{PQ})$ are given
as before by Eq.~(\ref{eqn:Psi1}) and Eq.~(\ref{eqn:Psi2}).  The
terms associated with the complex moduli in ${\xi}^{j,k}_{PQ}$ and
${\xi}^{j,k\bar{l}}_{PQ}$ are the same as those in
Eq.~(\ref{idb:eq:dthdu}) and Eq.~(\ref{idb:eq:dth2du}), and the
terms associated with the dilaton are given by
\begin{equation}
{\xi}^{j,s}_{PQ} \equiv (s+\bar s)\,\frac{\partial
\theta^j_{PQ}}{\partial s}=
 \left[-\frac{1}{4\pi}
 \sin(2\pi\theta^j),
 \right]^P_Q \label{idb:eq:dthdus2}
\end{equation}
\begin{equation}
{\xi}^{j,k\bar s}_{PQ} \equiv (u^k+\bar
u^k)(s+\bar s)\,\frac{\partial^2 \theta^j_{PQ}}{\partial
u^k\partial\bar s}= \left\{\begin{array}{cc}
  \frac{1}{16\pi}\left[\sin{4\pi\theta^j}\right]^P_Q & \mbox{when} \ j=k \vspace*{0.6cm} \\
  -\frac{1}{16\pi}\left[\sin{4\pi\theta^j}\right]^P_Q & \mbox{when} \ j\neq k, \vspace*{0.2cm}
\end{array}\right.\label{idb:eq:dth2duds}
\end{equation}
and
\begin{equation}
{\xi}^{j,s\bar s}_{PQ} \equiv (s+\bar
s)(s+\bar s)\,\frac{\partial^2 \theta^j_{PQ}}{\partial
s\partial\bar s}=
  \frac{1}{16\pi}\left[\sin{4\pi\theta^j} + 4\sin(2\pi\theta^j)\right]^P_Q,\label{idb:eq:dth2dss}
\end{equation}
where $k,l\neq s$. As before, the $\Theta_i$ parameters are
constrained as $\sum_{i=1}^3 \Theta_i^2 + \Theta_s^2 = 1$. In
order to simplify the analysis, we fix $\Theta_s$ while varying
$\Theta_3$, $\Theta_1$, and $\Theta_2$ subject to the unitary
condition. Since there is now another free parameter, the possible
parameter space is much larger than in the previous case.  Some
points in the parameter space which give the observed dark matter
density are shown in Table~\ref{intnumPS1} where we have fixed
$\Theta_s = 0.40$.
\begin{table}[h]
\footnotesize \caption{Supersymmetry breaking soft terms at the
string scale and resulting neutralino relic density for some
specific choices of goldstino angles, with $\Theta_s = 0.40$.}
\begin{center}
\begin{tabular}{|c|c|c|c|c|c|c|c|c|c|c|c|c|c|c|c|c|c|c|} \hline
                 $\Theta_1$&$\Theta_2$&$\Theta_3$&$M_{\tilde{G}}$~(GeV)&$M_{\tilde{W}}$~(GeV)&$M_{\tilde{B}}$~(GeV)&$m_H$~(GeV)&$m_L$~(GeV)&$m_R$~(GeV)&$A_t$~(GeV)& LSP & $\Omega h^2$ \\ \hline\hline
$-0.490$&$0.640$ & $0.592$ & $932$ & $207$ & $344$ & $688$ &    $832$ & $659$ & $154$ & $\tilde{B}$&    $0.112$ \\
$-0.450$&$-0.45$ & $0.771$ &    $181$ & $-736$ & $181$ & $327$ & $470$ & $447$ & $842$ & $\tilde{B}/\tilde{H}$ & $0.118$\\
$0.140$ &$0.870$ & $0.473$ & $971$ &    $407$  & $592$ & $815$ & $470$ & $562$ & $-732$ & $\tilde{B}$ & $0.114$\\
$0.380$ &$-0.45$ & $0.808$ & $218$ &  $-736$ & $649$ & $142$ & $564$ & $671$ & $-68$ & $\tilde{B}/\tilde{H}$ & $0.105$\\
$0.480$ &$-0.41$ & $0.776$ & $220$ & $-701$ & $682$ & $312$ & $692$ & $638$ & $-213$ & $\tilde{B}/\tilde{H}$ & $0.112$\\
$0.600$ &$-0.13$ & $0.789$ & $476$ & $-458$ & $856$ & $255$ & $765$ & $520$ & $-516$ & $\tilde{B}$   & $0.112$ \\
\hline\hline

\end{tabular}
\end{center}
\label{intnumPS1}
\end{table}

We exhibit the relic density as a function of $\Theta_1$ and
$\Theta_2$ for the particular case with $\Theta_s = 0.40$ and
tan~$\beta = 46$ in Fig~6.

\begin{figure}[ht]
    \centering
        \includegraphics{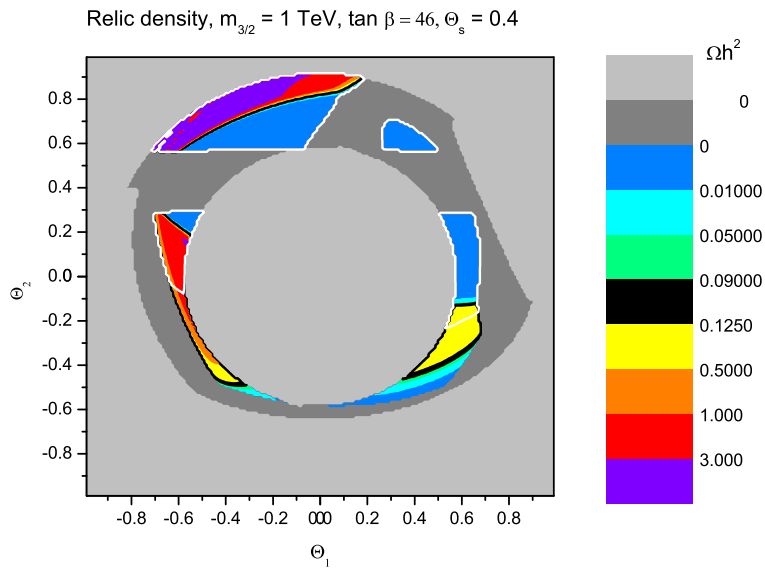}
    \caption{Contour maps of the relic density for tan~$\beta=46$,
    $\theta_s=0.40 \neq 0$, $\theta_3 = 0.75$, and $\mu > 0$ with a gravitino mass $m_{3/2} = 1$~TeV.
    The light gray regions are excluded because they do
    not satisfy the constraints on the soft terms at high scale.
    Regions inside the white contour satisfy the LEP limit, $m_h > 114$~GeV.}
    \label{fig:DMHiggs_1TeVS40}
\end{figure}

In the figure, only the regions within the white contour satisfy
the LEP limit on the Higgs mass.  Essentially, this rules out
regions of the parameter space with a mixed Bino/Higgsino LSP as
the Higgs mass for these regions of the parameter space is always
below $114$~GeV. A sampling of some of the superpartner spectra for some
of the cases of Table~\ref{intnumPS1} are shown in
Tables~\ref{SSpectrum3},~\ref{SSpectrum4}, and~\ref{SSpectrum5}
below.  The viable parameter space is quite large as can be seen from 
Fig.~7. 

\begin{table}[htb]

\footnotesize

\renewcommand{\arraystretch}{1.0}

\caption{Low energy supersymmetric particles and their masses (in
GeV) for $\Theta_1 = -0.600$, $\Theta_2 = -0.130$, $\Theta_3 =
0.789$, and $\Theta_s = 0.40$ with tan~$\beta=46$ and
$m_{3/2}=1$~TeV.}

\label{SSpectrum3}

\begin{center}

\begin{tabular}{|c|c|c|c|c|c|c|c|c|c|}\hline

$h^0$ & $H^0$ & $A^0$ & $H^{\pm}$ & ${\tilde g}$ & $\chi_1^{\pm}$
& $\chi_2^{\pm}$ & $\chi_1^{0}$ & $\chi_2^{0}$  \\ \hline

114.28  & 600.19   & 600.2    & 606.28   & 1135  & 390    & 774.8   & -363.2
&  -390.0      \\ \hline

$\chi_3^{0}$ & $\chi_4^{0}$ & ${\tilde t}_1$ & ${\tilde t}_2$ &
${\tilde u}_1/{\tilde c}_1$ & ${\tilde u}_2/{\tilde c}_2$ &
${\tilde b}_1$ & ${\tilde b}_2$ & \\ \hline

769.5  & -772.8  & 856.5  & 1133. & 1253. & 1105. & 985.4 & 1126. &\\ \hline

${\tilde d}_1/{\tilde s}_1$ & ${\tilde d}_2/{\tilde s}_2$ &
${\tilde \tau}_1$ & ${\tilde \tau}_2$ & ${\tilde \nu}_{\tau}$ &
${\tilde e}_1/{\tilde \mu}_1$ & ${\tilde e}_2/{\tilde \mu}_2$ &
${\tilde \nu}_e/{\tilde \nu}_{\mu}$ & \\ \hline

1255. & 1091. & 470.4 & 796.9 & 787.6 & 834.2 & 607.9 & 830.5 & \\

\hline

\end{tabular}

\end{center}

\end{table}

\begin{table}[htb]

\footnotesize

\renewcommand{\arraystretch}{1.0}

\caption{Low energy supersymmetric particles and their masses (in
GeV) for $\Theta_1 = -0.490$, $\Theta_2 = 0.640$, $\Theta_3 =
0.592$, and $\Theta_s = 0.40$ with tan~$\beta=46$ and
$m_{3/2}=1$~TeV.}

\label{SSpectrum4}

\begin{center}

\begin{tabular}{|c|c|c|c|c|c|c|c|c|c|}\hline

$h^0$ & $H^0$ & $A^0$ & $H^{\pm}$ & ${\tilde g}$ & $\chi_1^{\pm}$
& $\chi_2^{\pm}$ & $\chi_1^{0}$ & $\chi_2^{0}$  \\ \hline

117.82  & 885.60   & 885.61    & 889.63  & 2079 & 155.1   & 1110.   & 139.1
&  155.1      \\ \hline

$\chi_3^{0}$ & $\chi_4^{0}$ & ${\tilde t}_1$ & ${\tilde t}_2$ &
${\tilde u}_1/{\tilde c}_1$ & ${\tilde u}_2/{\tilde c}_2$ &
${\tilde b}_1$ & ${\tilde b}_2$ & \\ \hline

-1107  & 1109  & 1573.  & 1774. & 1961. & 1900. & 1731 & 1793. &\\ \hline

${\tilde d}_1/{\tilde s}_1$ & ${\tilde d}_2/{\tilde s}_2$ &
${\tilde \tau}_1$ & ${\tilde \tau}_2$ & ${\tilde \nu}_{\tau}$ &
${\tilde e}_1/{\tilde \mu}_1$ & ${\tilde e}_2/{\tilde \mu}_2$ &
${\tilde \nu}_e/{\tilde \nu}_{\mu}$ & \\ \hline

 1962. & 1900 & 486.1 & 793.6 & 775.9 & 838.1 & 669.4 & 834.5 &\\

\hline

\end{tabular}

\end{center}

\end{table}

\begin{table}[h]

\footnotesize

\renewcommand{\arraystretch}{1.0}

\caption{Low energy supersymmetric particles and their masses (in
GeV) for $\Theta_1 = 0.140$, $\Theta_2 = 0.870$, $\Theta_3 =
0.473$, and $\Theta_s = 0.40$ with tan~$\beta=46$ and
$m_{3/2}=1$~TeV.}

\label{SSpectrum5}

\begin{center}

\begin{tabular}{|c|c|c|c|c|c|c|c|c|c|}\hline

$h^0$ & $H^0$ & $A^0$ & $H^{\pm}$ & ${\tilde g}$ & $\chi_1^{\pm}$
& $\chi_2^{\pm}$ & $\chi_1^{0}$ & $\chi_2^{0}$  \\ \hline

118.02  & 911.27   & 911.17    & 915.16  & 2136 & 323.6   & 1138.   & 247.4
&  323.6      \\ \hline

$\chi_3^{0}$ & $\chi_4^{0}$ & ${\tilde t}_1$ & ${\tilde t}_2$ &
${\tilde u}_1/{\tilde c}_1$ & ${\tilde u}_2/{\tilde c}_2$ &
${\tilde b}_1$ & ${\tilde b}_2$ & \\ \hline

-1134  & 1137  & 1515.  & 1691. & 1911. & 1933. & 1639 & 1762. &\\ \hline

${\tilde d}_1/{\tilde s}_1$ & ${\tilde d}_2/{\tilde s}_2$ &
${\tilde \tau}_1$ & ${\tilde \tau}_2$ & ${\tilde \nu}_{\tau}$ &
${\tilde e}_1/{\tilde \mu}_1$ & ${\tilde e}_2/{\tilde \mu}_2$ &
${\tilde \nu}_e/{\tilde \nu}_{\mu}$ & \\ \hline

 1912. & 1931 & 267.6 & 517.4 & 425.9 & 540.1 & 602.7 & 534.4 &\\

\hline

\end{tabular}

\end{center}

\end{table}

As before, it can be seen that regions of the corresponding
parameter space correspond to cases where the the lightest
neutralino mass is very close to the either or both the lightest
chargino/next-to-lightest neutralino and light stau mass.  Thus, the observed dark matter
density is obtained close to the $\tilde{\chi}^0_1~\tilde{\tau}_1$
and/or $\tilde{\chi}^0_1~\tilde{\chi}^{\pm}/\tilde{\chi}^0_1~\tilde{\chi}^{0}_2$
coanhilation regimes. Similar results hold for other values of tan $\beta$.

\begin{figure}[htb]
	\centering
		\includegraphics[width=0.50\textwidth]{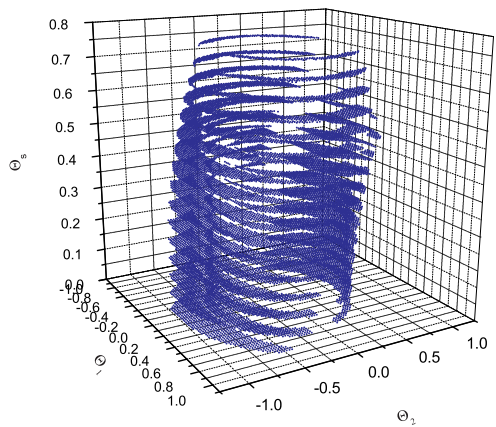}
	\label{fig:DMTotal}
	\caption{The viable parameter space with $\Omega h^2 \leq 0.125$ for $\Theta_s \geq 0$.} 
	\label{fig:DMTotal} 
\end{figure}

\section{Yukawa Couplings}

In addition to the fact that the SM fermions are replicated
into three distinct generations, the different generations
exhibit an intricate pattern of mass hierarchies.  At present,
there has been no satisfactory explanation for this. However, 
in addition to the replication of chirality, intersecting D-brane
models may naturally give rise to mass hierarchies and mixing.   

The Yukawa couplings in the intersecting D-brane worlds arise from
open string world-sheet instantons that connect three D-brane
intersections \cite{Aldazabal:2000cn}. For a given triplet
of intersections, the minimal action world-sheets which contribute
to the Yukawa coupling are weighted by a factor $exp(-A_{abc})$, where
$A_{abc}$ is the world-sheet area of the triangles bounded by the branes $a, b,$ and $c$. Since
there are several possible triangles with different areas, mass hierarchies may arise. One may also
see that
the Yukawa couplings depend on both the D-brane positions in the internal space, 
as well as on the geometry of the underlying compact manifold.  Effectively,
these quantities are parameterized by open and closed-string moduli VEVs.  Thus,
in order to make any definitive predictions for the Yukawa couplings, these moduli 
must be stabilized.  

Despite the promise of intersecting D-brane models in explaining the fermion mass hierarchies and mixings
of the Standard Model, they have typically been plagued by a rank-1 problem in the Yukawa mass matrices.  This can be traced to the fact that not all of the Standard Model fermions are localized at intersections on the same torus (in the case of toroidal orientifold compactifications). The general flavour structure and selection rules for intersecting D-brane
models as been investigated in~\cite{Chamoun:2003pf, Higaki:2005ie} However, the intersections are all on the first torus for the model of the previous section. Thus the resulting Yukawa mass matrices do not have a rank-1 problem.    In the following, we will explore the moduli space of this model in order
to see if there are any solutions which may give rise to realistic Yukawa textures,
following the analysis of 
\cite{Cremades:2003qj}, without first addressing the issue of moduli stabilization.  Our goal
for the present work is simply to see if the observed fermion mass hierarchies may arise
in this model and identify the points in the moduli space where this may happen.  We discuss
how the moduli might be fixed in a later section.

\subsection{The general form}

We start by considering D6 branes in Type IIA, where
the D6-branes wrap 3-cycles which intersect at angles on a
compact manifold $\mathbf{T^6}= \mathbf{T^2} \times \mathbf{T^2}
\times \mathbf{T^2}$.  For simplicity, we consider three stacks of
D-branes wrapping on a two-torus.  The 3-cycles wrapped by the D-branes can be
written in a vector form in terms of the wrapping numbers:
\begin{eqnarray}
&& [\Pi_a]=n_a [a] + m_a [b] : \; z_a = R(n + U m_a) \cdot x_a,
\nonumber \\
&& [\Pi_b]=n_b [a] + m_b [b] : \; z_b = R(n_b + U m_b) \cdot x_b,
\nonumber \\
&& [\Pi_c]=n_c [a] + m_c [b] : \; z_c = R(n_c + U m_c) \cdot x_c,
\end{eqnarray}
where $U$ is the complex structure parameter of the torus and $x\in
\mathbf{R}$ arbitrary numbers.  The Yukawa coupling involving
branes $a$, $b$, and $c$ recieves a contribution from the areas of the
triangles bounded by the triplet of D-branes.  To ensure that the triplet
of intersections actually form a 
triangle, we must impose the \textit{closer} condition
\cite{Cremades:2003qj}, 
\begin{equation}
z_a+z_b+z_c = 0. \label{close}
\end{equation}
The wrapping numbers are all integers due to the quantization
conditions, so by translating Eq. (\ref{close}) into the
Diophantine equation, the solution is found to be
\begin{eqnarray}
&&x_a = \frac{I_{bc}}{d} x, \nonumber \\
&&x_b = \frac{I_{ca}}{d} x, \;\; x=x_0+l, \;\; x_0 \in \mathbf{R},
\; l \in \mathbf{Z}, \nonumber \\
&&x_c = \frac{I_{ab}}{d} x,
\end{eqnarray}
where $I_{ab}$ is the intersection number, and $d=g.c.d.(I_{ab},
I_{bc}, I_{ca})$ is the greatest common divisor of the intersection
numbers.  The parameter $l$ indexes the different points in the
covering space $\mathbf{C}$ but the same points in the lattice of
$\mathbf{T^2}$ of the triangles.  The quantity $x_0$ is dependent on the
different intersection points of two of the D-branes which are
indexed by
\begin{eqnarray}
i=0,1,\cdots, (|I_{ab}|-1), \nonumber \\
j=0,1,\cdots, (|I_{bc}|-1), \nonumber \\
k=0,1,\cdots, (|I_{ca}|-1),
\end{eqnarray}
therefore $x_0$ can be written as
\begin{equation}
x_0(i,j,k) = \frac{i}{I_{ab}}+ \frac{j}{I_{ca}} +
\frac{k}{I_{bc}}.
\end{equation}

It is not necessary to require that all branes intersect at the
origin.  If the position of the stacks is shifted  by an amount
$\epsilon_{\alpha}$, $\alpha=a,b,c$, in clockwise directions of
stack $\alpha$ by a length in units of $A/||\Pi_{\alpha}||$ on
each torus, then $x_0$ will be modified as
\begin{equation}
x_0(i,j,k) = \frac{i}{I_{ab}}+ \frac{j}{I_{ca}} + \frac{k}{I_{bc}}
+ \frac{d ( I_{ab}\epsilon_c + I_{ca}\epsilon_b + I_{ab}\epsilon_a
)}{I_{ab} I_{bc} I_{ca}}. \label{x0shift}
\end{equation}
With this parameterizion of $x_0$, we can now calculate the areas of
the triangles, by the area formula of vectors
\begin{eqnarray}
&& A(z_a, z_b) = \frac{1}{2} \sqrt{|z_a|^2|z_b|^2-({\rm Re} z_a
\bar{z}_b)^2} \nonumber \\
\rightarrow && A_{ijk}(l) = \frac{1}{2} (2\pi)^2 A |I_{ab} I_{bc}
I_{ca}| (\frac{i}{I_{ab}} + \frac{j}{I_{ca}} + \frac{k}{I_{bc}} +
\epsilon + l)^2,
\end{eqnarray}
where $\epsilon$ is the total shift effect in Eq. (\ref{x0shift}),
\begin{equation}
\epsilon=\frac{I_{ab} \epsilon_c + I_{ca} \epsilon_b + I_{bc}
\epsilon_a}{I_{ab} I_{bc} I_{ca}}
\end{equation}
and $A$ is the K\"ahler structure of the torus, which is generally
the area.  By adding a real phase $\sigma_{abc} = {\rm
sign}(I_{ab} I_{bc} I_{ca})$ for the full instanton contribution,
the corresponding Yukawa coupling for the three states
localized at the intersections indexed by $(i,j,k)$ is
\cite{Cremades:2003qj}
\begin{equation}
Y_{ijk}= h_{qu} \sigma_{abc} \sum_{l \in Z}
\exp(-\frac{A_{ijk}(l)}{2 \pi \alpha'}),
\end{equation}
where $h_{qu}$ is due to quantum correction as discussed
in \cite{Cvetic:2003ch}.   The summation can be expressed in terms
of a modular theta function for convenient
numerical calculation. The real version of this theta function can
be written as
\begin{equation}
\vartheta \left[\begin{array}{c} \delta \\ \phi
\end{array} \right] (t) = \sum_{l\in\mathbf{Z}} e^{-\pi t (\delta+l)^2}
e^{2l\pi i(\delta + l) \phi}, \label{Rtheta}
\end{equation}
so after comparing the parameters, we have
\begin{eqnarray}
&&\delta = \frac{i}{I_{ab}} + \frac{j}{I_{ca}} + \frac{k}{I_{bc}}
+ \epsilon, \\
&& \phi=0, \\
&& t=\frac{A}{\alpha'} |I_{ab} I_{bc} I_{ca}|.
\end{eqnarray}

\paragraph{B-field and Wilson lines \\ \\}

The theta function above is
constrained by its real property.  However, $t$ can be complex and
$\phi$ can be any number as an overall phase which can be given both a
theoretical and phenomenological interpretation.

If we turn on a B-field in the compact space $\mathbf{T^2}$, the
string will not only couple to the metric but also to this
B-field.  Then the K\"ahler structure may be written in a complex form
\begin{equation}
J=B+iA,
\end{equation}
where the parameter $t$ is replaced by a complex parameter $\kappa$
\begin{equation}
\kappa = \frac{J}{\alpha'} |I_{ab} I_{bc} I_{ca}|.
\end{equation}

We can also include Wilson lines around the compact directions
that the D-branes wrap in this construction \cite{Cremades:2003qj}.
To avoid breaking any gauge symmetry, these Wilson lines are
chosen up to a phase. If we consider the Yukawa coupling formed by
D-branes $a$, $b$, and $c$, wrapping a different one-cycle with
Wilson lines in the phases $\exp (2\pi i \theta_a)$, $\exp (2\pi i
\theta_b)$, and $\exp (2\pi i \theta_c)$ respectively, then the
total phase is a linear combination of each phase weighted by the
relative longitude of each segment, determined by the intersection
points:
\begin{equation}
e^{2\pi i x_a \theta_a}  e^{2\pi i x_b \theta_b} e^{2\pi i x_c
\theta_c} = e^{2\pi i x(I_{bc} \theta_a + I_{ca} \theta_b + I_{ab}
\theta_c)}.
\end{equation}

Thus, including these two effects, we obtain a general complex theta
function as 
\begin{equation}
\vartheta \left[\begin{array}{c} \delta \\ \phi
\end{array} \right] (\kappa) = \sum_{l\in\mathbf{Z}} e^{\pi i \kappa
(\delta+l)^2 } e^{2 \pi i(\delta + l) \phi}, \label{Ctheta}
\end{equation}
where
\begin{eqnarray}
&&\delta = \frac{i}{I_{ab}} + \frac{j}{I_{ca}} + \frac{k}{I_{bc}}
+ \epsilon, \\
&& \phi=I_{bc} \theta_a + I_{ca} \theta_b +
I_{ab} \theta_c, \\
&& \kappa=\frac{J}{\alpha'} |I_{ab} I_{bc} I_{ca}|.
\end{eqnarray}

\paragraph{Other modification \\ \\}

In most of the (semi-)realistic models orientifold planes must be
introduced to cancel the RR-tadpoles.  As a result,
there will be additional fields from the brane images coupling to
the ordinary branes fields as well as themselves.  For example, for
the triangle formed by branes $a$, $b'$, and $c$, the Yukawa
coupling will then depend on the parameters $I_{ab'}$, $I_{b'c}$, and
$I_{ca}$, and the corresponding indicies are $i'$, $j$, and
$k'$, where the primed indexes are independent of the unprimed
ones.

The other issue is the non-coprime cases.  The three intersection
numbers are not necessarily coprime, so we have to make sure we do
not over-count the repeated parts.   The constant $d$ is defined as the 
$g.c.d.$ of
the intersection numbers and is introduced in the brane shift
parameters. We must then modify the other parameters as well:
\begin{eqnarray}
&&  \phi = \frac{I_{bc} \theta_a + I_{ca} \theta_b +
I_{ab} \theta_c}{d}, \\
&&  \kappa=\frac{J}{\alpha'} \frac{|I_{ab} I_{bc} I_{ca}|}{d^2}.
\end{eqnarray}

There is one more constraint which is necessary to ensure non-zero
Yukawa couplings: the triangles must be bounded by D-branes. 
Thus the intersection indexes $i$, $j$, and $k$ cannot
be arbitrary. They are required to satisfy \cite{Cremades:2003qj}
\begin{equation}
i+j+k = 0 \ {\rm~mod} \ d.  \label{closetriA}
\end{equation}
There is one degree of freedom which may ease this constraint, which
is an additional parameter in $\delta$: \cite{Cremades:2003qj}
\begin{equation}
\delta = \frac{i}{I_{ab}} + \frac{j}{I_{ca}} + \frac{k}{I_{bc}} +
\epsilon + \frac{s}{d},
\end{equation}
where $s$ is a linear combination of $i$, $j$, and $k$, and it is just a
shift of counting the triangles since we have required $\{i,j,k\}$
to satisfy (\ref{closetriA}).

Therefore finally, we can write down a complete form for the
Yukawa couplings for D6-branes wrapping on a full
compact space $\mathbf{T^2} \times \mathbf{T^2} \times
\mathbf{T^2}$ as
\begin{equation}
Y_{\{ijk\}}=h_{qu} \sigma_{abc} \prod_{r=1}^3 \vartheta
\left[\begin{array}{c} \delta^{(r)}\\ \phi^{(r)}
\end{array} \right] (\kappa^{(r)}),
\end{equation}
where
\begin{equation}
\vartheta \left[\begin{array}{c} \delta^{(r)}\\ \phi^{(r)}
\end{array} \right] (\kappa^{(r)})=\sum_{l_r\in\mathbf{Z}} e^{\pi
i(\delta^{(r)}+l_r)^2 \kappa^{(r)}} e^{2\pi i(\delta^{(r)}+l_r)
\phi^{(r)}},   \label{Dtheta}
\end{equation}
with $r=1,2,3$ denoting the three two-tori. The input parameters
are given by
\begin{eqnarray} 
\nonumber
&&\delta^{(r)} = \frac{i^{(r)}}{I_{ab}^{(r)}} +
\frac{j^{(r)}}{I_{ca}^{(r)}} + \frac{k^{(r)}}{I_{bc}^{(r)}} +
\frac{d^{(r)} ( I_{ab}^{(r)} \epsilon_c^{(r)} + I_{ca}^{(r)}
\epsilon_b^{(r)} + I_{ab}^{(r)} \epsilon_a^{(r)}
)}{I_{ab} I_{bc} I_{ca}} + \frac{s^{(r)}}{d^{(r)}}, \\ \nonumber
&&\phi^{(r)} = \frac{I_{bc}^{(r)} \theta_a^{(r)} + I_{ca}^{(r)}
\theta_b^{(r)} + I_{ab}^{(r)} \theta_c^{(r)}}{d^{(r)}}, \\ 
&&\kappa^{(r)} = \frac{J^{(r)}}{\alpha'} \frac{|I_{ab}^{(r)}
I_{bc}^{(r)} I_{ca}^{(r)}|}{(d^{(r)})^2}.
\label{eqn:Yinput}
\end{eqnarray}

\paragraph{Theta function with characters \\ \\}

It can be complicated to calculate the numerical value of
the theta function defined in Eq. (\ref{Ctheta}).  Thus, for simplicity
the B-field will not be introduced in the following analysis.  Then
if we define $J'= -iJ = A$ and so $\kappa'= -i\kappa$ for
convenience, the theta function turns out to be
\begin{eqnarray}
&&\vartheta \left[\begin{array}{c} \delta \\ \phi
\end{array} \right] (\kappa') = \sum_{l\in\mathbf{Z}} e^{-\pi
\kappa' (\delta+l)^2 } e^{2 \pi i(\delta + l) \phi}, \nonumber\\
\stackrel{\textrm{redefine}}{\longrightarrow} &&\vartheta
\left[\begin{array}{c} \delta \\ \phi \end{array} \right] (\kappa)
= e^{-\pi \kappa \delta^2} e^{2\pi i \delta \phi} \vartheta_3 (\pi
(\phi+ i \kappa \delta), e^{-\pi \kappa}), \label{Ntheta}
\end{eqnarray}
where $\vartheta_3$ is the Jacobi theta function of the third kind.
A plot of the theta function is shown in Fig.~\ref{fig:Thetafunction},
with $\phi=0$. 

\begin{figure}[ht]
	\centering
		\includegraphics{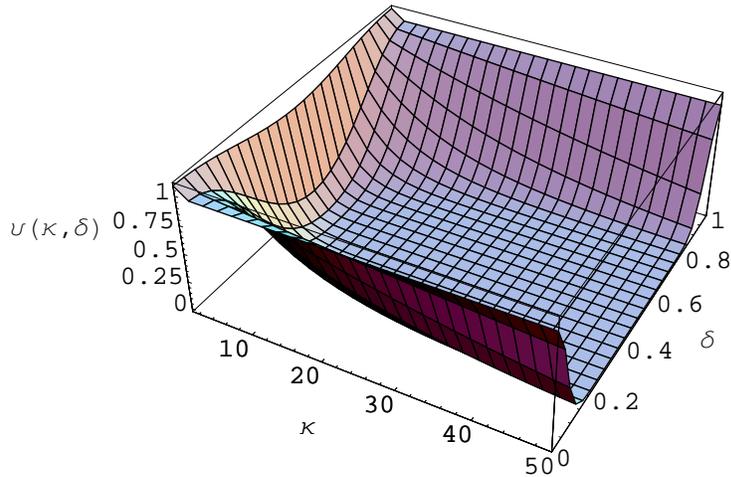}
	\caption{Theta function as a function of the parameters $\delta$ and $\kappa$, with $\phi=0$. The maximum
	values occur for $\delta = 0$ while the minimum values occur when $\delta = 1/2$.}
	\label{fig:Thetafunction}
\end{figure}

\subsection{Semi-Realistic Yukawa Textures}

(i) Mass Matrices

As mentioned previously, the intersecting D-brane model with matter content shown in Table
\ref{MI-Numbers} has several desirable semi-realistic features,
namely three-generations of chiral SM fermions with a minimum of exotic matter, tree-level
gauge coupling unification, and the fact that the three intersections
required to form the disk diagrams for the Yukawa couplings all occur on
the first torus as can be seen from  Figure \ref{3tori}.  Thus, in our
analysis we will focus on the just the first torus, since the
contribution from the
other two tori will just give an over-all constant. This constant,which is
different for the up-type and down-type quark, and charged lepton
mass matrices, is unimportant for the present analysis since we will only
obtain the mass ratios rather than the absolute fermion masses. 

As described in the previous analysis, the Pati-Salam gauge symmetry is broken
to the Standard Model by a process which involves brane-splitting,
\begin{equation}
a \rightarrow \ a_1 + a_2, \ \ \ \ \ c \rightarrow c_1 + c_2.
\end{equation}
so that the Standard Model quarks and leptons arise from
\begin{eqnarray}
F_L(Q_L, L_L)  \rightarrow   Q_L + L \nonumber \\
F_R(Q_R, L_R)  \rightarrow   U_R + D_R + E_R + N.
\end{eqnarray}
The Yukawa couplings for the quarks and leptons are then given by the superpotential
\begin{equation}
W_Y = Y^U_{ijk} Q_L^i U_R^j H_U^k + Y^D_{ijk} Q_L^i D_R^j H_D^k + Y^L_{ijk} L^i E^j H_D^k, 
\end{equation}
where it should be kept in mind that there are six vector pairs of Higgs multiplets, each of
which may receive a VEV.  

\begin{figure}[h]
\begin{center}
\includegraphics[width=.8\textwidth,angle=0]{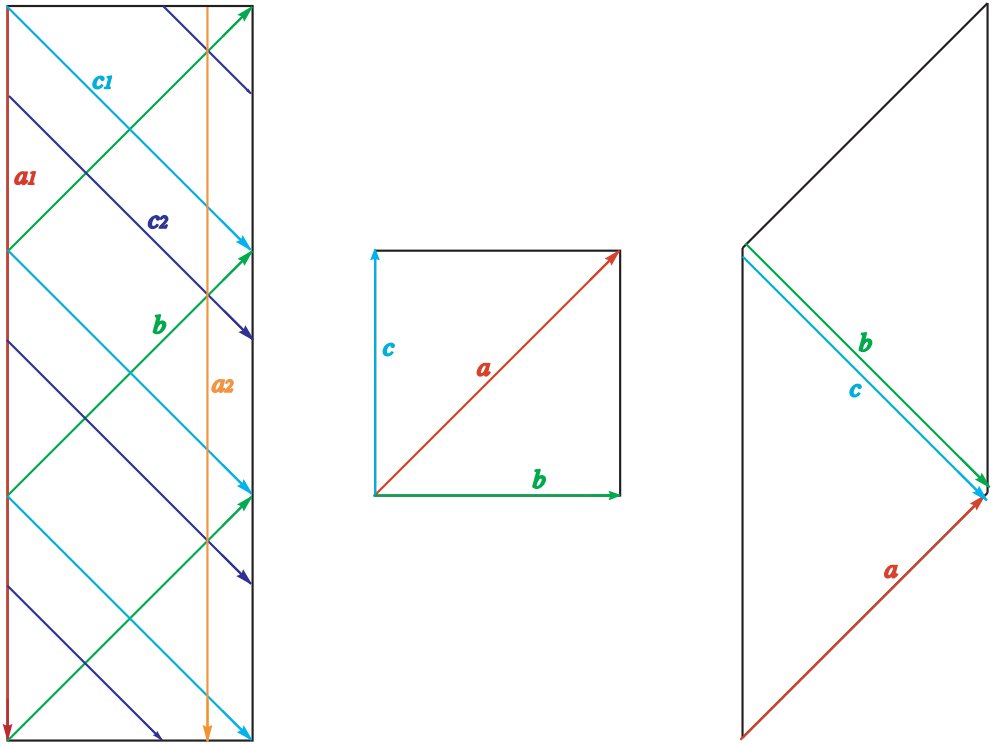}
\caption{Brane configuration for the three two-tori.  The SM fermion mass
hierarchies primarily result from the intersections on the first torus.}  \label{3tori}
\end{center}
\end{figure}

For the model under consideration, the intersection numbers on each torus are
given by
\begin{center}
\begin{tabular}{l l l}
\\
$I_{ab}^{(1)}=3$, & $I_{ab}^{(2)}=-1$, & $I_{ab}^{(3)}=-1$, \\
$I_{ca}^{(1)}=-3$, & $I_{ca}^{(2)}=-1$, & $I_{ca}^{(3)}=1$,  \\
$I_{bc}^{(1)}=-6$, & $I_{bc}^{(2)}=1$, & $I_{bc}^{(3)}=0$. \\
\end{tabular}
\end{center}
These intersection numbers are not coprime, so that we need to have
$d^{(1)}=g.c.d.(I_{ab}^{(1)},I_{bc}^{(1)},I_{ca}^{(1)})=3$,
$d^{(2)}=1$.  Note that we do not need $d^{(3)}$ since 
the intersections do not form any triangles on the third torus.  
Thus the parameters of the theta
functions defined in Eq.~\ref{eqn:Yinput} are given by
\begin{eqnarray}
&&\delta^{(1)} = \frac{i^{(1)}}{3} - \frac{j^{(1)}}{3} -
\frac{k^{(1)}}{6} + \frac{\epsilon_c^{(1)} -
\epsilon_b^{(1)} - 2\epsilon_a^{(1)}}{6} + \frac{s^{(1)}}{3},  \nonumber  \\
&&\delta^{(2)} = -\epsilon_c^{(2)} - \epsilon_b^{(2)} +
\epsilon_a^{(2)},    \nonumber  \\
&&\delta^{(3)} = -\epsilon_c^{(3)} + \epsilon_b^{(3)},
\\ \nonumber \\
&&\phi^{(1)} =\;\;\; \theta_c^{(1)} - \theta_b^{(1)}
- 2\theta_a^{(1)}, \nonumber \\
&&\phi^{(2)} =-\theta_c^{(2)} - \theta_b^{(2)} + \theta_a^{(2)},
\nonumber \\
&&\phi^{(3)} =-\theta_c^{(3)} + \theta_b^{(3)},  \\ \nonumber \\
&&\kappa^{(1)} = \frac{6J^{(1)}}{\alpha'}, \nonumber \\
&&\kappa^{(2)} = \frac{J^{(2)}}{\alpha'}, \nonumber \\
&&\kappa^{(3)} = 0,
\end{eqnarray}
where $i=0\dots 2$, $j=0\dots 2$, and $k=0\dots 5$, indexing the
left-handed fermions, right-handed fermions, and Higgs fields respectively.
For mathematical convenience
we  redefine the shift on each torus as
\begin{eqnarray}
&&\epsilon^{(1)} \equiv \frac{\epsilon_c^{(1)} - \epsilon_b^{(1)}
- 2\epsilon_a^{(1)}}{6}, \nonumber \\
&&\epsilon^{(2)} \equiv -\epsilon_c^{(2)} - \epsilon_b^{(2)} +
\epsilon_a^{(2)},  \nonumber \\
&&\epsilon^{(2)} \equiv -\epsilon_c^{(3)} + \epsilon_b^{(3)}.
\label{shifts}
\end{eqnarray}

The intersection numbers on  the second and third tori are either one or zero,
so their contribution to the Yukawa couplings will be an over-all constant.   
For a given triplet of intersections to be
connected by an instanton, the selection rule
\begin{equation}
i^{(1)} + j^{(1)} + k^{(1)} = 0\; \mathrm{ mod }\; 3
\end{equation}
should be satisfied.  Then, the Yukawa coupling
matrices will take the following form:

\begin{eqnarray}
Y^{(1)}_{k=0}   \sim  \left(\begin{array}{@{}c@{}@{}c@{}@{}c@{}}
a_{000} & 0 & 0 \\
0 & 0 & a_{120} \\
0 & a_{210} & 0  \end{array} \right),                 \;\;
Y^{(1)}_{k=1}   \sim  \left(\begin{array}{@{}c@{}@{}c@{}@{}c@{}}
0 & 0 & a_{021} \\
0 & a_{111} & 0 \\
a_{201} & 0 & 0  \end{array} \right),                 \;\;
Y^{(1)}_{k=2}   \sim  \left(\begin{array}{@{}c@{}@{}c@{}@{}c@{}}
0 & a_{012} & 0 \\
a_{102} & 0 & 0 \\
0 & 0 & a_{222}  \end{array} \right),      \nonumber \\     \;\;
Y^{(1)}_{k=3} \sim \left(\begin{array}{@{}c@{}@{}c@{}@{}c@{}}
a_{003} & 0 & 0 \\
0 & 0 & a_{123} \\
0 & a_{213} & 0  \end{array} \right),                 \;\;
Y^{(1)}_{k=4}   \sim  \left(\begin{array}{@{}c@{}@{}c@{}@{}c@{}}
0 & 0 & a_{024} \\
0 & a_{114} & 0 \\
a_{204} & 0 & 0  \end{array} \right),                 \;\;
Y^{(1)}_{k=5}   \sim  \left(\begin{array}{@{}c@{}@{}c@{}@{}c@{}}
0 & a_{015} & 0 \\
a_{105} & 0 & 0 \\
0 & 0 & a_{225}  \end{array} \right).
\end{eqnarray}

By choosing a different linear function for $s^{(1)}$, some independent
modes with non-zero eigenvalues are available, which are listed
below.

\textbf{(i)} $s^{(1)}=0$
\begin{eqnarray}
&&a_{000}=a_{102}=a_{204}=\vartheta \left[\begin{array}{c}
\epsilon^{(1)}\\ \phi^{(1)} \end{array} \right]
(\frac{6J^{(1)}}{\alpha'}) \equiv A,  \nonumber \\
&&a_{210}=a_{012}=a_{114}=\vartheta \left[\begin{array}{c}
\epsilon^{(1)}+\frac{1}{3}\\  \phi^{(1)} \end{array} \right]
(\frac{6J^{(1)}}{\alpha'}) \equiv B,  \nonumber \\
&&a_{120}=a_{222}=a_{024}=\vartheta \left[\begin{array}{c}
\epsilon^{(1)}-\frac{1}{3}\\  \phi^{(1)} \end{array} \right]
(\frac{6J^{(1)}}{\alpha'}) \equiv C, \nonumber   \\
&&a_{021}=a_{123}=a_{225}=\vartheta \left[\begin{array}{c}
\epsilon^{(1)}+\frac{1}{6}\\  \phi^{(1)} \end{array} \right]
(\frac{6J^{(1)}}{\alpha'}) \equiv D, \nonumber   \\
&&a_{201}=a_{003}=a_{105}=\vartheta \left[\begin{array}{c}
\epsilon^{(1)}+\frac{1}{2}\\  \phi^{(1)} \end{array} \right]
(\frac{6J^{(1)}}{\alpha'}) \equiv E, \nonumber   \\
&&a_{111}=a_{213}=a_{015}=\vartheta \left[\begin{array}{c}
\epsilon^{(1)}-\frac{1}{6}\\  \phi^{(1)} \end{array} \right]
(\frac{6J^{(1)}}{\alpha'}) \equiv F,
\end{eqnarray}
in other words
\begin{eqnarray}
Y^{(1)}_{k=0}   \sim  \left(\begin{array}{ccc}
A & 0 & 0 \\
0 & 0 & C \\
0 & B & 0  \end{array} \right),                      \;\;
Y^{(1)}_{k=1} \sim  \left(\begin{array}{ccc}
0 & 0 & D \\
0 & F & 0 \\
E & 0 & 0  \end{array} \right),                      \;\;
Y^{(1)}_{k=2} \sim  \left(\begin{array}{ccc}
0 & B & 0 \\
A & 0 & 0 \\
0 & 0 & C  \end{array} \right),   \nonumber \\
Y^{(1)}_{k=3}   \sim  \left(\begin{array}{ccc}
E & 0 & 0 \\
0 & 0 & D \\
0 & F & 0  \end{array} \right),                      \;\;
Y^{(1)}_{k=4} \sim  \left(\begin{array}{ccc}
0 & 0 & C \\
0 & B & 0 \\
A & 0 & 0  \end{array} \right),                      \;\;
Y^{(1)}_{k=5} \sim  \left(\begin{array}{ccc}
0 & F & 0 \\
E & 0 & 0 \\
0 & 0 & D  \end{array} \right).
\end{eqnarray}

\textbf{(ii)} $s^{(1)}=j$

\begin{eqnarray}
Y^{(1)}_{k=0}   \sim  \left(\begin{array}{ccc}
A & 0 & 0 \\
0 & 0 & B \\
0 & C & 0  \end{array} \right),                      \;\;
Y^{(1)}_{k=1} \sim  \left(\begin{array}{ccc}
0 & 0 & F \\
0 & D & 0 \\
E & 0 & 0  \end{array} \right),                      \;\;
Y^{(1)}_{k=2} \sim  \left(\begin{array}{ccc}
0 & C & 0 \\
B & 0 & 0 \\
0 & 0 & A  \end{array} \right),   \nonumber \\
Y^{(1)}_{k=3}   \sim  \left(\begin{array}{ccc}
E & 0 & 0 \\
0 & 0 & F \\
0 & D & 0  \end{array} \right),                      \;\;
Y^{(1)}_{k=4} \sim  \left(\begin{array}{ccc}
0 & 0 & B \\
0 & C & 0 \\
A & 0 & 0  \end{array} \right),                      \;\;
Y^{(1)}_{k=5} \sim  \left(\begin{array}{ccc}
0 & D & 0 \\
F & 0 & 0 \\
0 & 0 & E  \end{array} \right).
\end{eqnarray}

\textbf{(iii)} $s^{(1)}=-i$

\begin{eqnarray}
Y^{(1)}_{k=0}   \sim  \left(\begin{array}{ccc}
A & 0 & 0 \\
0 & 0 & B \\
0 & C & 0  \end{array} \right),                      \;\;
Y^{(1)}_{k=1} \sim  \left(\begin{array}{ccc}
0 & 0 & D \\
0 & E & 0 \\
F & 0 & 0  \end{array} \right),                      \;\;
Y^{(1)}_{k=2} \sim  \left(\begin{array}{ccc}
0 & B & 0 \\
C & 0 & 0 \\
0 & 0 & A  \end{array} \right),   \nonumber \\
Y^{(1)}_{k=3}   \sim  \left(\begin{array}{ccc}
E & 0 & 0 \\
0 & 0 & F \\
0 & D & 0  \end{array} \right),                      \;\;
Y^{(1)}_{k=4} \sim  \left(\begin{array}{ccc}
0 & 0 & C \\
0 & A & 0 \\
B & 0 & 0  \end{array} \right),                      \;\;
Y^{(1)}_{k=5} \sim  \left(\begin{array}{ccc}
0 & F & 0 \\
D & 0 & 0 \\
0 & 0 & E  \end{array} \right).
\end{eqnarray}

The cases $s^{(1)}=k/2$ and $s^{(1)}=2j$ are not considered since
there they may forbid three different real
eigenvalues.  We will take case (iii) for the
following discussion.

The most general form for a given mass matrix can then be given as
\begin{equation}
\mathcal{M}   \sim  \left(\begin{array}{ccc}
A v_1 + E v_4 & B v_3 + F v_6 & D v_2 + C v_5 \\
C v_3 + D v_6 & A v_5 + E v_2 & B v_1 + F v_4 \\
F v_2 + B v_5 & C v_1 + D v_4 & A v_3 + E v_6 \end{array} \right),
\label{Yukawa general}
\end{equation}
where $v_i=\langle H_i\rangle$ are the different VEVs of the six Higgs
fields present in the model.
\begin{equation}
(M_u)_{ij}   \sim \left(\begin{array}{ccc}
A^U H^1_u + E^U H^4_u & B^U H^3_u + F^U H^6_u & D^U H^2_u + C^U H^5_u \\
C^U H^3_u + D^U H^6_u & A^U H^5_u + E^U H^2_u & B^U H^1_u + F^U H^4_u \\
F^U H^2_u + B^U H^5_u & C^U H^1_u + D^U H^4_u & A^U H^3_u + E^U
H^6_u
\end{array} \right), \label{Yukawa generalU}
\end{equation}
\begin{equation}
(M_d)_{ij}   \sim \left(\begin{array}{ccc}
A^D H_d^1 + E^D H_d^4 & B^D H_d^3 + F^D H_d^6 & D^D H_d^2 + C^D H_d^5 \\
C^D H_d^3 + D^D H_d^6 & A^D H_d^5 + E^D H_d^2 & B^D H_d^1 + F^D H_d^4 \\
F^D H_d^2 + B^D H_d^5 & C^D H_d^1 + D^D H_d^4 & A^D H_d^3 + E^D
H_d^6
\end{array} \right), \label{Yukawa generalD}
\end{equation}
\begin{equation}
(M_l)_{ij}   \sim \left(\begin{array}{ccc}
A^E H_d^1 + E^E H_d^4 & B^E H_d^3 + F^E H_d^6 & D^E H_d^2 + C^E H_d^5 \\
C^E H_d^3 + D^E H_d^6 & A^E H_d^5 + E^E H_d^2 & B^E H_d^1 + F^E H_d^4 \\
F^E H_d^2 + B^E H_d^5 & C^E H_d^1 + D^E H_d^4 & A^E H_d^3 + E^E
H_d^6
\end{array} \right). \label{Yukawa generalL}
\end{equation}

The two light Higgs mass eigenstates which arise by fine-tuning the superpotential
shown in Eq.~\ref{eqn:HiggsSup} would then be different linear combinations of
the six Higgs fields present in the model.  For a given set of VEVs for
the Higgs fields chosen to obtain the desired Yukawa matrices, we can then write the
light Higgs eigenstates as
\begin{equation}
H_{u,d}= \sum \frac{v^i_{u,d}}{\sqrt{\sum (v^i_{u,d})^2}} H^i_{u,d},
\label{Higgseig}
\end{equation}
with $v^i_{u,d} = \langle H^i_{u,d}\rangle$.
 
One may see from the form of the above mass matrices that there is a natural
mass hierarchy among the up-quarks, down-quarks
and leptons which has a geometric interpretation related
to the position of the different D-branes, as parameterized by the 
shifts $\epsilon_i$. Recall that the Pati-Salam gauge symmetry has
been broken to the SM by a process which involves splitting the stacks of
D-branes as shown in Fig.~\ref{brnsplit}. The left-handed quarks are localized
at the intersections between stacks $a1$ and $b$, the right-handed up-type quarks
are localized between stacks $a1$ and $c1$, while the right-handed down-type quarks
are localized between stacks $a1$ and $c2$.  Thus, if stack $c2$ is shifted on the
torus by an amount $\epsilon_{c2}$ while stack $c1$ is unshifted ($\epsilon_{c1} = 0$),
then the down-type quark masses are naturally suppressed relative to the up-type quarks.
Similarly, the left-handed charged leptons are localized at the intersection 
between stack $a2$ and $b$, while the right-handed charged leptons are localized
at the intersection between stacks $a2$ and $c2$.  Since stack $a2$ will be shifted
by some amount $\epsilon_{a2}$, the resulting charged lepton masses will be
naturally suppressed relative to the down-type quarks.  Thus, from purely
geometric considerations, we expect 
\begin{equation}
m_{u} > m_{d} > m_{l}.  
\end{equation}  

It is also clear from the form of the mass matrices that the mass hierarchies
between the different up-type quarks among themselves, the different down-type
quarks among themselves, and the different charged leptons among themselves
are due primarily to the different Higgs neutral component VEVs.  
Effectively, these values will determined by
fine-tuning the superpotential so that there are only two Higgs eigenstates
as shown in Eq.~\ref{Higgseig}.  Ideally,  
one would like to be able to dynamically determine the Higgs
eigenstates from first principles rather than fine-tuning the superpotential. 
However, it does not seem possible to do this at least until the issue
of moduli stablization has been fully addressed.   
   
If we define
$D_u$ and $D_d$ as the mass diagonal matrices of up and down-type
quarks respectively,
\begin{equation}
U^u_L M_u {U^u_R}^{\dag}=D_u,\;\; U^d_L M_d {U^d_R}^{\dag}=D_d,
\;\; V_{CKM}=U^u_L {U^d_L}^{\dag},
\end{equation}
where $U^i$ are unitary matrices, then the squared mass matrices are $M_u M_u^{\dag}$ and $M_d
M_d^{\dag}$.
In the Standard Model, we can always make quark mass matrices $M_u$
and $M_d$ Hermitian by suitable transformation of the right-handed
fields.  If we take a case that the $M_d$ is very close to the
diagonal matrix for down-type quark, in other words, $U^d_L$ and
$U^d_R$ are very close to the unit matrix with very small
off-diagonal terms, then
\begin{equation}
V_{CKM}\simeq U^u U^{d\dag}\simeq U^u,
\end{equation}
where we have transformed away the right-handed effects and make
them the same as the left-handed ones.  Then the mass matrix of
the up-type quarks is
\begin{equation}
M_u \sim V_{CKM}^{\dag} D_u V_{CKM}.
\end{equation}

 (ii) An Example
 
For different superpartner spectra, we may determine the mass matrices for quarks 
and leptons 
by running the RGE's up to the unification scale.  For example, for tan$\beta \approx 50$
the CKM matrix at the 
unification scale $\mu=M_X$ may be determined to be~\cite{Fusaoka:1998vc,Ross:2007az}

\begin{equation}
V_{CKM} = \left(\begin{array}{ccc}
0.9754 & 0.2205 & -0.0026i  \\
-0.2203e^{0.003^{\circ}i} & 0.9749 & 0.0318 \\
0.0075e^{-19^{\circ}i} & -0.0311e^{1.0^{\circ}i} & 0.9995
\end{array} \right),
\end{equation}
and the diagonal quark mass matrices $D_u$ and $D_d$ may be written
\begin{equation}
D_u = m_t \left(\begin{array}{ccc}
0.0000139 & 0 & 0  \\
0 & 0.00404 & 0 \\
0 & 0 & 1
\end{array} \right), \;\;
D_d = m_b \left(\begin{array}{ccc}
0.00141 & 0 & 0  \\
0 & 0.0280 & 0 \\
0 & 0 & 1
\end{array} \right).
\end{equation}
Then the absolute value of $M_u$ turns out to be
\begin{equation}
|M_u| = m_t \left(\begin{array}{ccc}
0.000266 & 0.00109 & 0.00747  \\
0.00109 & 0.00481 & 0.0310 \\
0.00747 & 0.0310 & 0.999
\end{array} \right).
\label{Umass}
\end{equation}

We can then fine-tune the parameters and
Higgs VEVs in Eqs. (\ref{Yukawa generalU}), (\ref{Yukawa generalD}), and
(\ref{Yukawa generalL}) to fit these mass matrices. It looks at the first glance that the solution can be easily
found, but we should keep in mind that the six parameters from the
theta function controlled by the D-brane shifts and Wilson-line
phases are \textit{not} independent. In doing so, we are 
essentially constrained by the off-diagnonal terms, as can be seen by
comparing Eq.~\ref{Yukawa generalU} with Eq.~\ref{Umass}.  For example,
the $(11)$ element of 
$|M_u|$ is $0.00109$, while the corresponding element of the fitted
mass matrix is $B v_3 + F v_6$.  However, the diagonal $(33)$ element of the fitted
matrix is given by $A v_3 + E v_6$ which should be of order unity.  Thus, to fit
the mass matrix 
we must require 
\begin{equation}
\frac{B v_3 + F v_6}{A v_3 + E v_6} \approx 0.00109, 
\label{Offdiagonalcon}
\end{equation}
which effectively means that $\frac{B}{A} << 1$ and $\frac{F}{E} << 1$. 

\begin{figure}
	\centering
		\includegraphics{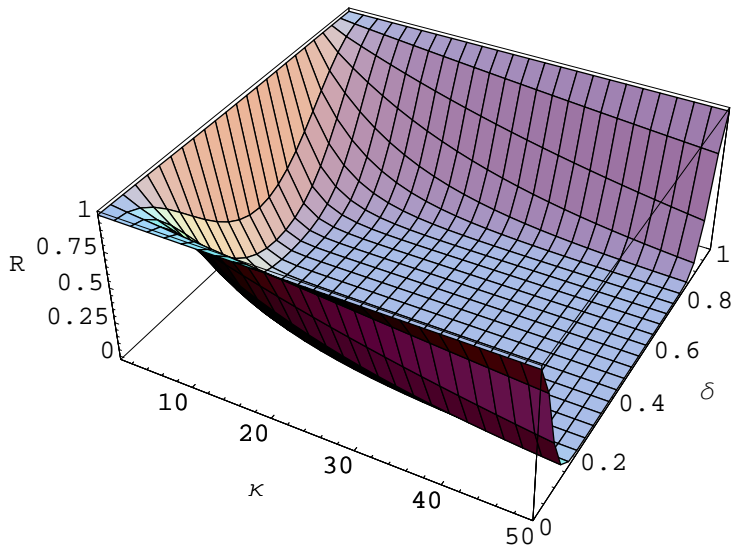}
	\caption{Ratio of theta functions $R=[e^{-\pi \kappa \delta^2} \vartheta_3 (i  \pi
\kappa \delta, e^{-\pi \kappa})]/[\vartheta_3 (0, e^{-\pi \kappa})]$.}        
					
	\label{fig:Thetaratio}
\end{figure}

In Fig.~\ref{fig:Thetaratio}, we plot the ratio 
\begin{equation}
\frac{\vartheta
\left[\begin{array}{c} \delta \\ 0 \end{array} \right] (\kappa)
= e^{-\pi \kappa \delta^2} \vartheta_3 (i  \pi
\kappa \delta, e^{-\pi \kappa})}{\vartheta
\left[\begin{array}{c} \delta=0 \\ 0 \end{array} \right] (\kappa)
=  \vartheta_3 (0, e^{-\pi \kappa})}.
\label{ratio}
\end{equation}
It can be seen from this plot that the off-diagonal terms will be naturally suppressed relative to
the diagonal elements.  In order to satisfy the conditions shown in Eq.~\ref{Offdiagonalcon} to fit
the up-type quark mass matrix,
must must allow $\kappa$ to be large enough such that $e^{-\pi \kappa}\approx 0$.  This in effect
means that $\vartheta_3 (i  \pi
\kappa \delta, e^{-\pi \kappa}) = 1$, so that  
\begin{eqnarray}
A = e^{-\pi\kappa\epsilon^2}, \ \ \ B = e^{-\pi\kappa (\epsilon+1/3)^2}, \ \ \ C = e^{-\pi\kappa (\epsilon-1/3)^2}, \nonumber \\
D = e^{-\pi\kappa (\epsilon+1/6)^2}, \ \ \ E = e^{-\pi\kappa (\epsilon+1/2)^2}, \ \ \ F = e^{-\pi\kappa (\epsilon-1/6)^2}.
\end{eqnarray}

On comparing $M_u$ and Eq. (\ref{Yukawa generalU}) we can obtain a set of
parameters which
will give the desired solution.  For
example, if we set $\frac{6J^{(1)}}{\alpha'}= \kappa = 39.6$ and
$\epsilon^{(1)U} = 0$, where $\epsilon^{(1)U}_a =
\epsilon^{(1)}_{a^1} = 0$, $\epsilon^{(1)U}_b = \epsilon^{(1)}_b =
0$, and $\epsilon^{(1)U}_c = \epsilon^{(1)}_{c^1} = 0$, then
\begin{equation}
\begin{array}{l c l}
A^U=1,        &&  v^1_u= 0.000266,   \\
B^U=0.000001, &&  v^2_u= 0.236,     \\
C^U=0.000001, &&  v^3_u= 0.999,      \\
D^U=0.0316,   &&  v^4_u= 0.981,      \\
E^U=0.0,      &&  v^5_u= 0.00481,    \\
F^U=0.0316,   &&  v^6_u= 0.0345,
\end{array}
\end{equation}
which may reproduce exactly the mass matrix Eq.~\ref{Umass}.  

Similarly, after fixing the K\"ahler
modulus from what is needed to obtain the
correct off-diagonal terms for the up-type quark mass matrix,
we can then obtain a result for $M_d$.
For the present case, we have $\frac{m_{\tau}}{m_b}=1.58$ which is
obtained from the previous analysis of the soft-terms for a point
in the parameter space satisfying all constraints. With the choices
$\epsilon^{(1)D}=0.061$, where $\epsilon^{(1)D}_a =
\epsilon^{(1)}_{a^1} = 0$, $\epsilon^{(1)D}_b = \epsilon^{(1)}_b =
0$, and $\epsilon^{(1)D}_c = \epsilon^{(1)}_{c^2} = 0.366$ (by
Eq. (\ref{shifts})), the desired solution may be obtained with
\begin{equation}
\begin{array}{l c l}
A^D=0.629,    &&  v_d^1= 0.00224,   \\%
B^D=0.0,      &&  v_d^2= 0.0,     \\
C^D=0.000098, &&  v_d^3= 1.58,      \\%
D^D=0.00158,  &&  v_d^4= 0.0,      \\
E^D=0.0,      &&  v_d^5= 0.0445,    \\%
F^D=0.249,    &&  v_d^6= 0.0001,
\end{array}
\end{equation}
The calculated mass matrix $M_d$ is thus given as
\begin{equation}
|M_d| \sim m_b \left(\begin{array}{ccc}
0.00141 & 0.000025 & 0.000004  \\
0.000155 & 0.0280 & 0.0 \\
0.0 & 0.000000220 & 1
\end{array} \right) \sim D_d.
\end{equation}

Note that the down-type quark mass matrix and the lepton mass matrix both involve the same Higgs fields.  Thus, once the parameters needed to fit the down-type mass matrix are fixed, the only freedom in calculating the lepton mass matrix is from the geometric position of each brane.  To fit the lepton matrix consistent with the down-type quark matrix, we may choose $\epsilon^{(1)L}=0.0$, where
$\epsilon^{(1)L}_a = \epsilon^{(1)}_{a^2} = 0.183$,
$\epsilon^{(1)L}_b = \epsilon^{(1)}_b = 0.0$, $\epsilon^{(1)L}_c =
\epsilon^{(1)}_{c^2} = 0.366$, then
\begin{equation}
\begin{array}{l c l}
A^L=1.0,     &&  B^L=0.000001,     \\
C^L=0.000001,&&  D^L=0.0316,         \\
E^L=0.0,     &&  F^L=0.0316,
\end{array}
\end{equation}
so that the fitted mass matrix of leptons is given as
\begin{eqnarray}
|M_l| \sim m_b \left(\begin{array}{ccc}
0.00224 & 4.74\times 10^{-6} & 4.45\times 10^{-8}  \\
4.74\times 10^{-6} & 0.0445 & 2.24\times 10^{-9} \\
4.45\times 10^{-8} & 2.24\times 10^{-9} & 1.58
\end{array} \right) \\= m_{\tau}\left(\begin{array}{ccc}
0.00142 & 3.0\times 10^{-6} & 2.82\times 10^{-8}  \\
3.0\times 10^{-6} & 0.0282 & 1.42\times 10^{-9} \\
2.82\times 10^{-8} & 1.42\times 10^{-9} & 1.0
\end{array} \right).
\end{eqnarray}
We may compare the eigenvalues $m_{\tau} \{0.014, 0.028, 1\}$ for the fitted matrix with
the extrapolated lepton masses obtained from running the RGEs up to the GUT scale:  
\begin{equation}
D_l = m_{\tau} \left(\begin{array}{ccc}
0.000217 & 0 & 0  \\
0 & 0.0458 & 0 \\
0 & 0 & 1
\end{array} \right).
\end{equation}
Thus, the electron mass comes out to be too big by a factor of six, while
the muon mass is 60\% too small.  However, the tau lepton mass does come out
correctly.  This result can be understood by considering that the only difference
between the down-type quark and lepton mass matrices is an overall exponential factor
which results from the additional shift for the leptonic brane, $\epsilon_{a1}$.  
Although this result seems to give the wrong answers for the electron
and muon masses, it should be kept in mind that these are only tree-level results.  Indeed, it
is of interest that the error in the obtained lepton masses seems to increase with decreasing mass.  
There could indeed be other corrections, such as those coming from higher-dimensional operators, which
would contribute most greatly to the electron and muon masses since they are quite small.

In short, the above mass matrices can produce the correct quark masses and
CKM mixings, and the correct $\tau$ lepton mass at the electroweak
scale. The electron mass is about $5\sim6.5$ times larger than the
expected value, while the muon mass is about $50\sim60\%$ too small.
Similar to GUTs, we end up with roughly the wrong fermion mass
relation $m_e/m_{\mu} \cong m_d/m_s$. The correct electron and
muon masses may in principle be generated via high-dimensional operators by 
introducing the vector-like Higgs fields from the $ac'$ sector~\cite{INPREP}.
Moreover, the suitable neutrino masses and mixings can be
generated via the seesaw mechanism by choosing suitable Majorana
mass matrix for the right-handed neutrinos.

\section{A Comment on Moduli Stabilization}

In the previous section, it has been demonstrated that it possible 
to obtain Yukawa textures which can reproduce the observed fermion
mass hierarchies and mixings (extrapolated at the unification scale). 
Although this is a very interesting result, it is
clear from this analysis that the Yukawa couplings depend on several 
parameters which are not determined within the model.  Rather, random
values for these parameters have been chosen by hand in order to fit
the experimental results.  Thus, it is far from clear that this model in 
its current form can offer a completely satisfactory explanation for the observed fermion
mass hierarchies and mixings.  

Given a concrete string model, the low-energy observables such as particle
couplings and resulting masses are functions of the open and closed string
moduli.  In particular, we can see from the analysis of the previous section
that the Yukawa couplings depend on the position of each stack of branes on 
the tori as well as on the K\a"ahler moduli.  Thus, if these moduli can
be fixed dynamically, then the possible Yukawa mass textures would be 
tightly constrained.  Indeed, once these moduli are fixed, the only remaining
freedom is in the Higgs sector, namely the specific linear combination of 
the six pairs of Higgs states which must be fine-tuned to produce the two
Higgs eigenstates $H_u$ and $H_d$ of the MSSM.  

The job of fixing the position of each stack of branes on each torus is 
equivalent to fixing the open-string moduli.  D-brane constructions typically 
have non-chiral open string states present in the low-energy spectrum associated 
with the D-brane position in the internal space and Wilson lines.  
This results in adjoint or additional matter in the
symmetric and antisymmetric representations unless the open 
string moduli are completely frozen.   These light scalars are not observed 
and the succesful gauge unification in the MSSM would also be spoiled by their presence.  
While it may be possible to find some scenarios where the problems created by these 
fields are ameliorated, it is much simpler to eliminate these fields altogether.  
One way to do this is to this is to construct intersecting D-brane models where 
the D-branes wrap rigid cycles, which was first explored 
in~\cite{Dudas:2005jx} and~\cite{Blumenhagen:2005tn} in the
context of Type II compactifications on $T^6/(\Z_2 \times \Z_2')$ which is 
the only known toroidal background which posseses such rigid cycles. 

For $T^6/(\Z_2 \times \Z_2')$ the twisted homology contains collapsed 3-cycles. 
There are 16 fixed points, from which arise 16 additional 2-cycles with the topology of $\mathbf{P}^1 \cong S^2$.  As a result, there are 32 collapsed 3-cycles for each twisted sector.  A $D6$-brane wrapping collapsed 3-cycles in each of the three twisted sectors will be unable to move away from a particular position on the covering space $\mathbf{T^6}$, and thus the 3-cycle will be rigid.   

A fractional D-brane wrapping both a bulk cycle as well as the collapsed cycles may be written in the form
\begin{eqnarray}
\Pi^F_a &=& \frac{1}{4}\Pi^B + \frac{1}{4}\left(\sum_{i,j\in S^a_{\theta}} \epsilon^{\theta}_{a,ij}\Pi^{\theta}_{ij,a}\right)+ \frac{1}{4}\left(\sum_{j,k\in S^a_{\omega}} \epsilon^{\omega}_{a,jk}\Pi^{\omega}_{jk,a}\right)
+ \frac{1}{4}\left(\sum_{i,k\in S^a_{\theta\omega}} \epsilon^{\theta\omega}_{a,ik}\Pi^{\theta\omega}_{ik,a}\right).
\label{fraccycle}
\end{eqnarray}
where the $D6$-brane is required to run through the four fixed points for each of the twisted sectors.  The set of four fixed points may be denoted as $S^g$ for the twisted sector $g$. The constants $\epsilon^{\theta}_{a,ij}$, $\epsilon^{\omega}_{a,jk}$ and $\epsilon^{\theta\omega}_{a,ki}$ denote the sign of the charge of the fractional brane with respect to the fields which are present at the orbifold fixed points.  We refer the reader to \cite{Blumenhagen:2005tn} for a detailed discussion of model building on this background.  

Let us consider a local supersymmetric model consisting of three stacks of D6 branes wrapping fractional cycles with the bulk wrapping numbers and intersection numbers shown in Table~\ref{rigidmodel}.  This model is essentially equivalent to the model we have been studying, modulo the fact that the D6 branes are now wrapping fractional cycles
and there are no longer any flat directions for the open-string moduli.  Thus, the positions of the D-branes are frozen
since they are unable to move away from the fixed points.  Phenomenologically, this is very desirable.  First, the existence of the adjoint which results from unstablized open-string moduli can destroy the gauge-coupling unification and the asymptotic freedom of $SU(3)_C$.  Second, fixing the open-string moduli means that there is limited freedom
in the values taken by the brane-shift parameters ($\epsilon_i$ of the previous section). As we have seen, these parameters play a fundamental role in generating the mass hierarchies between up-type and down-type quarks, and the leptons.  In addition, the open-string moduli must be fixed in order to calculate parameters such as $\mu$ and tan $\beta$ which play a very important
role in determining the low-energy superpartner spectra and the Yukawa couplings. Finally, instanton induced superpotential couplings 
may also be very important for addressing issues such as neutrino masses, inflation and supersymmetry breaking. The Euclidean D-branes necessary for calculating such effects must wrap rigid
cycles.  

\begin{table}[t]
\footnotesize
\renewcommand{\arraystretch}{1.0}
\caption{D6-brane configurations and bulk intersection numbers for
a local left-right model in Type IIA on the $\mathbf{T}^6 /(\Z_2 \times \Z_2')$
orientifold, where the D6-branes are wrapping rigid cycles.  Although the bulk intersection numbers are the same for stacks $a_1$
and $a_2$, it should be understood that the
cycles wrapped by these stacks go through different fixed points on at least one torus.}
\label{rigidmodel}
\begin{center}
\begin{tabular}{|c||c|c||c|c|c|c|c|c|c|c|c|c|c|c|c|}
\hline
& \multicolumn{11}{c|}{$SU(3)_C\times SU(2)_L\times SU(2)_R \times U(1)_{B-L}$}\\
\hline \hline  & $N$ & $(n^1,l^1)\times (n^2,l^2)\times

(n^3,l^3)$ & $n_{S}$& $n_{A}$ & $b$ & $b'$ &  $c$ & $c'$ &$\epsilon^{\theta}_{ij}~\forall~ij$ &  $\epsilon^{\omega}_{jk}~\forall~jk$ & $\epsilon^{\theta\omega}_{ki}~\forall~kl$\\

\hline

    $a_1$&  3 & $(1,0)\times (1,-1)\times (1,1)$ & 0 & 0  & -3 & 1&  3 & -1 & -1 & -1 & 1\\
    
    $a_2$&  1 & $(1,0)\times (1,-1)\times (1,1)$ & 0 & 0  & -3 & 1&  3 & -1 & -1 & -1 & 1\\

    $b$&  2 & $(1,3)\times (1,0)\times (1,-1)$ & 0 & 4  & - & - & 0 & 2  & 1 & 1 & 1\\

    $c$&  2& $(1,-3)\times (0,1)\times (1,-1)$ & -8 & 0  & - & -  & - & - & 1 & 1 & 1 \\

\hline
\end{tabular}

\end{center}

\end{table}

In order to have such rigid cycles, we must make a choice of discrete torsion which is related to the sign
of the orientifold planes.  Namely, in order for the background to have discrete torsion, there must be
an odd number of $O6^{(+,+)}$ planes.  Thus, the conditions necessary for tadpole cancellation depend directly
on the choice of discrete torsion, which will then determine what combination of hidden sector
branes and/or flux is necessary in order to have a globally consistent model.  The hidden sector will also be necessary to cancel twisted-tadpoles associated with the orbifold fixed points.   

Of course, in the end, our strategy will be to construct the model in the T-dual
picure involving magnetized fractional D-branes, in the context of flux compactifications.  Indeed, it is necessary to 
turn on fluxes in order to stabilize the closed-string moduli.  In fact, one such model with an equivalent observable
sector as the one we have been studying has been constructed as examples of supersymmetric Type IIA Ads and Type IIB Minkowski flux vacua~\cite{Chen:2006gd, Chen:2007af}. The next step would then be to combine these two sources of moduli stablization into a single construction with both open and closed-string moduli stabilized.  Hopefully, we would then be able to uniquely calculate both the Yukawa couplings as well as the superpartner spectra.  We are presently working on this, and hope
to report on our progress in the near future.

\newpage
\section{Conclusion}

We have analyzed in detail a three-family intersecting D6-brane
model where gauge coupling unification is achieved at the string
scale and where the gauge symmetry can be broken to the Standard
Model. In the model, it is possible to calculate the supersymmetry
breaking soft terms and obtain the low energy supersymmetric
particle spectra within the reach of the LHC. Finally, it is
possible to obtain the SM quark masses and CKM mixings and the 
lepton masses, and the neutrino masses and mixings may be generated
via the seesaw mechanism.

Clearly, this model cannot be regarded as being fully 
realistic until the moduli stablization
issue has been fully addressed. There are many free parameters
which have been fixed in order to obtain the desired values for
the Yukawa mass matrices and the value of the gauge couplings at
the unification scale, although it should be kept in mind that
these parameters are tightly constrained and it is not possible to
tune them to just any value. In the case of the Yukawa matrices,
the free parameters are the K\a"ahler moduli, the brane positions
on each torus (open-string moduli) and the specific linear
combination of states with which have been indentified the two
pairs of Higgs eigenstates. Although we have chosen specific
values for the moduli fields to obtain agreement with experiments,
it may be possible to uniquely predict these values by introducing
the most general fluxes.  It might also be possible to fix the
open string moduli if we require the D-branes wrap rigid cycles. 
However, it seems likely that the Higgs eigenstates would still
need to be fine-tuned.

On the other hand, this does appear to be the first such
string-derived model where it is possible to give mass to each
family of quarks and leptons.  Even if we cannot at present {\it
uniquely} predict these values, it must still be regarded as
highly significant in that it is {\it possible} to come very close to
getting correct mass matrices and mixings at the unification scale.
This suggests that the model may be a
candidate for a phenomenological description of elementary
particle physics in much the same way as the MSSM.  

It is also very appealing that the tree-level gauge couplings are unified at
the string scale, although it is still an open question if the
running of the gauge couplings can be maintained all the way down
to the electroweak scale.  The reason for this is that there are 
chiral exotic states present in the spectrum which are bifundamentals
under the observable and hidden sector gauge groups. We should note that 
most of these chiral
exotic states can be decoupled at the string scale and the rest may
may be decoupled at an intermediate scale.  Even if this were
not the case, we have found 
that the hidden sector
gauge interactions will become confining at around $10^{7}$~GeV
and $10^{13}$~GeV respectively, and so states charged under these
groups will not be present in the low-energy spectrum.  However,
there are exotic states in the spectrum which transform as
representations of both the hidden and observable sectors which
may effect the RGE running of the gauge couplings, although we do
not expect them to effect the running that much.  If we are
optimistic, then it is possible that these would amount to
threshold corrections which might push the unification scale up to
the string scale.

If the model does turn out to be a realistic effective description
of the observed elementary particle physics, then it should be
possible to predict the low-energy superpartner mass spectra from
the model.  Besides the D-brane wrapping numbers and closed-string
moduli, the superpartner spectra depend strongly on the exact way
in which supersymmetry is broken. In principle, it should be
possible to completely specify the exact mechanism, whether
through gaugino condensation in the hidden sector, flux-induced
soft terms, or via instanton induced couplings. In the present 
analysis, we have studied
supersymmetry breaking generically via a generic
parameterized F-term and have shown that it is possible to
constrain the phenomenologically allowed parameter space by
imposing experimental limits on the neutralino relic density and
mass limits coming from LEP. We have found that the viable parameter 
space is quite large.  Once the experimentally determined
superpartner mass spectrum begins to take shape it may be possible
to find a choice of F-terms which will correspond to the
observed spectrum.  It would be very interesting to explore 
the collider signatures which this model may produce at LHC for the
regions of the parameter space which satisfy all phenomenological 
constraints.

In summary, the model we have studied may produce a realistic
phenomenology, once the issue of moduli stabilization has been
fully addressed.  The model
represents the first known intersecting D-brane model for which mass may
be given to each generation of quarks and leptons.  Furthermore,
the supersymmetry breaking soft terms may be studied in the model
and may yield realistic superpartner spectra.  Certainly, the
model and the current theoretical tools are not presently
developed to the point where specific predictions for the known
fermion masses and superpartner spectrum may be made.  However, it
is clearly possible for the model to describe the known
physics of the Standard Model, as well as potentially describing new physics.

\section*{Acknowledgements}

This research was supported in part by the Mitchell-Heep Chair in
High Energy Physics (CMC), by the Cambridge-Mitchell Collaboration
in Theoretical Cosmology and by the Chinese Academy of Sciences under
 Grant KJCX3-SYW-N2(TL), and by the DOE grant
DE-FG03-95-Er-40917.


\newpage
\begin{thebibliography}{99}

\itemsep 0.5mm


\bibitem{JPEW}
  J.~Polchinski and E.~Witten,
  Nucl.\ Phys.\  B {\bf 460}, 525 (1996)
  [arXiv:hep-th/9510169].


\bibitem{bdl}
  M.~Berkooz, M.~R.~Douglas and R.~G.~Leigh,
  Nucl.\ Phys.\  B {\bf 480}, 265 (1996)

  [arXiv:hep-th/9606139].


\bibitem{bachas}
  C.~Bachas,
  [arXiv:hep-th/9503030].

\bibitem{Witten9810188}
  E.~Witten,
  JHEP {\bf 9812}, 019 (1998).

\bibitem{Blumenhagen:2000wh}
  R.~Blumenhagen, L.~G\"orlich, B.~K\"ors and D.~L\"ust,
  JHEP {\bf 0010}, 006 (2000)
  [arXiv:hep-th/0007024].


\bibitem{Angelantonj:2000hi}
  C.~Angelantonj, I.~Antoniadis, E.~Dudas and A.~Sagnotti,
  Phys.\ Lett.\ B {\bf 489}, 223 (2000)
  [arXiv:hep-th/0007090].



\bibitem{Blumenhagen:2005mu}
  R.~Blumenhagen, M.~Cveti\v c, P.~Langacker and G.~Shiu,
  Ann.\ Rev.\ Nucl.\ Part.\ Sci.\  {\bf 55}, 71 (2005)
  [arXiv:hep-th/0502005],
  and the references therein.




\bibitem{CSU1}
M.~Cveti\v c, G.~Shiu and A.~M.~Uranga, Phys.\ Rev.\ Lett.\  {\bf
87}, 201801 (2001).

\bibitem{CSU2}
M.~Cveti\v c, G.~Shiu and A.~M.~Uranga, Nucl.\ Phys.\ B {\bf 615},
3 (2001).


\bibitem{Cvetic:2002pj}
  M.~Cveti\v c, I.~Papadimitriou and G.~Shiu,
  Nucl.\ Phys.\ B {\bf 659}, 193 (2003)
  [Erratum-ibid.\ B {\bf 696}, 298 (2004)].


\bibitem{CP} M. Cveti\v c and I. Papadimitriou,
Phys.\ Rev.\ D {\bf 67}, 126006 (2003).


\bibitem{CLL}
M.~Cveti\v c, T.~Li and T.~Liu,
Nucl.\ Phys.\ B {\bf 698}, 163 (2004).

\bibitem{Cvetic:2004nk}
  M.~Cveti\v c, P.~Langacker, T.~Li and T.~Liu,
  Nucl.\ Phys.\ B {\bf 709}, 241 (2005).


\bibitem{Chen:2005ab}
  C.-M.~Chen, G.~V.~Kraniotis, V.~E.~Mayes, D.~V.~Nanopoulos and J.~W.~Walker,
  Phys.\ Lett.\ B {\bf 611}, 156 (2005);
Phys.\ Lett.\  B {\bf 625}, 96 (2005).


\bibitem{Chen:2005mj}
  C.-M.~Chen, T.~Li and D.~V.~Nanopoulos,
  Nucl.\ Phys.\ B {\bf 732}, 224 (2006).

\bibitem{Dudas:2005jx}
  E.~Dudas and C.~Timirgaziu,
  Nucl.\ Phys.\  B {\bf 716}, 65 (2005)
  [arXiv:hep-th/0502085].

\bibitem{Blumenhagen:2005tn}
  R.~Blumenhagen, M.~Cvetic, F.~Marchesano and G.~Shiu,
  JHEP {\bf 0503}, 050 (2005)
  [arXiv:hep-th/0502095].

\bibitem{Chen:2006sd}
  C.~M.~Chen, V.~E.~Mayes and D.~V.~Nanopoulos,
  Phys.\ Lett.\  B {\bf 648}, 301 (2007)
  [arXiv:hep-th/0612087].


\bibitem{CLS1}
 M.~Cveti\v c, P.~Langacker and G.~Shiu,
Phys.\ Rev.\ D {\bf 66}, 066004 (2002);
 Nucl.\ Phys.\ B {\bf 642}, 139 (2002).


\bibitem{CLW}
M.~Cveti\v c, P.~Langacker and J.~Wang,
Phys.\ Rev.\ D {\bf 68}, 046002 (2003).


\bibitem{ListSUSYOthers}
R.~Blumenhagen, L.~G\"orlich and T.~Ott, JHEP {\bf 0301}, 021
(2003);
G.~Honecker, Nucl.\ Phys.\  {\bf B666}, 175 (2003);
G.~Honecker and T.~Ott,
Phys.\ Rev.\ D {\bf 70}, 126010 (2004)
  [Erratum-ibid.\ D {\bf 71}, 069902 (2005)].

\bibitem{Blumenhagen:2006xt}
  R.~Blumenhagen, M.~Cvetic and T.~Weigand,
  arXiv:hep-th/0609191.

\bibitem{Ibanez:2006da}
  L.~E.~Ibanez and A.~M.~Uranga,
  JHEP {\bf 0703}, 052 (2007).


\bibitem{Cvetic:2007ku}
  M.~Cvetic, R.~Richter and T.~Weigand,
  arXiv:hep-th/0703028.

\bibitem{Ibanez:2007rs}
  L.~E.~Ibanez, A.~N.~Schellekens and A.~M.~Uranga,
  arXiv:0704.1079 [hep-th].


\bibitem{Kachru:Blumen}
  S.~Kachru, M.~B.~Schulz and S.~Trivedi,
  JHEP {\bf 0310}, 007 (2003)
  [arXiv:hep-th/0201028];
  R.~Blumenhagen, D.~L\"ust and T.~R.~Taylor,
  Nucl.\ Phys.\  B {\bf 663}, 319 (2003)
  [arXiv:hep-th/0303016].

\bibitem{Grimm:Zwirner}
  T.~W.~Grimm and J.~Louis,
  Nucl.\ Phys.\ B {\bf 718}, 153 (2005)
  [arXiv:hep-th/0412277];
  G.~Villadoro and F.~Zwirner,
  JHEP {\bf 0506}, 047 (2005)
  [arXiv:hep-th/0503169].

\bibitem{Cascales:2003zp}
  J.~F.~G.~Cascales and A.~M.~Uranga,
  JHEP {\bf 0305}, 011 (2003).


\bibitem{MS}
F.~Marchesano and G.~Shiu, Phys.\ Rev.\ D {\bf 71}, 011701 (2005);
JHEP {\bf 0411}, 041 (2004).

\bibitem{CL}
M.~Cveti\v c and T.~Liu, Phys.\ Lett.\ B {\bf 610}, 122 (2005).


\bibitem{Cvetic:2005bn}
  M.~Cveti\v c, T.~Li and T.~Liu,
  Phys.\ Rev.\ D {\bf 71}, 106008 (2005).


\bibitem{Kumar:2005hf}
  J.~Kumar and J.~D.~Wells,
  JHEP {\bf 0509}, 067 (2005).


\bibitem{Chen:2005cf}
  C.-M.~Chen, V.~E.~Mayes and D.~V.~Nanopoulos,
  Phys.\ Lett.\  B {\bf 633}, 618 (2006).

\bibitem{Blumenhagen:2006ci}
  R.~Blumenhagen, B.~Kors, D.~Lust and S.~Stieberger,
  arXiv:hep-th/0610327.

\bibitem{Camara:2005dc}
  P.~G.~Camara, A.~Font and L.~E.~Ibanez,
  JHEP {\bf 0509}, 013 (2005).


\bibitem{Chen:2006gd}
  C.-M.~Chen, T.~Li and D.~V.~Nanopoulos,
  Nucl.\ Phys.\  B {\bf 740}, 79 (2006).


\bibitem{Chen:2006ip}
  C.-M.~Chen, T.~Li and D.~V.~Nanopoulos,
  Nucl.\ Phys.\  B {\bf 751}, 260 (2006).

\bibitem{Chen:2007px}
  C.-M.~Chen, T.~Li, V.~E.~Mayes and D.~V.~Nanopoulos,
  arXiv:hep-th/0703280.

\bibitem{Chen:2007af}
  C.~M.~Chen, T.~Li, Y.~Liu and D.~V.~Nanopoulos,
  arXiv:0711.2679 [hep-th].

\bibitem{Bennett:2003bz}
  C.~L.~Bennett {\it et al.}  [WMAP Collaboration],
  Astrophys.\ J.\ Suppl.\  {\bf 148}, 1 (2003)
  [arXiv:astro-ph/0302207].

\bibitem{Spergel:2003cb}
  D.~N.~Spergel {\it et al.}  [WMAP Collaboration],
  Astrophys.\ J.\ Suppl.\  {\bf 148}, 175 (2003)
  [arXiv:astro-ph/0302209].


\bibitem{Blumenhagen:2001te}
  R.~Blumenhagen, B.~K\"{o}rs, D.~L\"{u}st and T.~Ott,
  Nucl.\ Phys.\  B {\bf 616}, 3 (2001)
  [arXiv:hep-th/0107138].

\bibitem{Cremades:2002te}
  D.~Cremades, L.~E.~Ib\'{a}\~{n}ez and F.~Marchesano,
  JHEP {\bf 0207}, 009 (2002)
  [arXiv:hep-th/0201205].


\bibitem{Shiu:1998pa}
  G.~Shiu and S.~H.~H.~Tye,
  Phys.\ Rev.\  D {\bf 58}, 106007 (1998)
  [arXiv:hep-th/9805157].


\bibitem{Lust:2003ky}
  D.~L\"ust and S.~Stieberger,
  [arXiv:hep-th/0302221].


\bibitem{Antoniadis:Blumen}
  I.~Antoniadis, E.~Kiritsis and T.~N.~Tomaras,
  Phys.\ Lett.\  B {\bf 486}, 186 (2000)
  [arXiv:hep-ph/0004214];
  R.~Blumenhagen, D.~Lust and S.~Stieberger,
  JHEP {\bf 0307}, 036 (2003)
  [arXiv:hep-th/0305146].

\bibitem{Cremades:2003qj}
  D.~Cremades, L.~E.~Ib\'{a}\~{n}ez and F.~Marchesano,
  JHEP {\bf 0307}, 038 (2003)
  [arXiv:hep-th/0302105].

\bibitem{Cvetic:2003ch}
  M.~Cveti\v{c} and I.~Papadimitriou,
  Phys.\ Rev.\  D {\bf 68}, 046001 (2003)
  [Erratum-ibid.\  D {\bf 70}, 029903 (2004)]
  [arXiv:hep-th/0303083].

\bibitem{Kors:2003wf}
  B.~K\"ors and P.~Nath,
  Nucl.\ Phys.\  B {\bf 681}, 77 (2004)
  [arXiv:hep-th/0309167].

\bibitem{Lust:2004cx}
  D.~L\"ust, P.~Mayr, R.~Richter and S.~Stieberger,
  Nucl.\ Phys.\  B {\bf 696}, 205 (2004)
  [arXiv:hep-th/0404134].

\bibitem{Kawamura:1996ex}
  Y.~Kawamura, T.~Kobayashi and T.~Komatsu,
  Phys.\ Lett.\  B {\bf 400}, 284 (1997)
  [arXiv:hep-ph/9609462].

\bibitem{Brignole:1993dj}
  A.~Brignole, L.~E.~Ibanez and C.~Munoz,
  Nucl.\ Phys.\  B {\bf 422}, 125 (1994)
  [Erratum-ibid.\  B {\bf 436}, 747 (1995)]
  [arXiv:hep-ph/9308271];
  [arXiv:hep-ph/9707209].

\bibitem{Blumenhagen:2003jy}
  R.~Blumenhagen, D.~Lust and S.~Stieberger,
  JHEP {\bf 0307}, 036 (2003)
  [arXiv:hep-th/0305146].

\bibitem{Ellis:1990iu}
  J.~R.~Ellis, J.~L.~Lopez and D.~V.~Nanopoulos,
  Phys.\ Lett.\  B {\bf 247}, 257 (1990).

\bibitem{Benakli:1998ut}
  K.~Benakli, J.~R.~Ellis and D.~V.~Nanopoulos,
  Phys.\ Rev.\  D {\bf 59}, 047301 (1999)
  [arXiv:hep-ph/9803333].

\bibitem{Ellis:2004cj}
  J.~R.~Ellis, V.~E.~Mayes and D.~V.~Nanopoulos,
  Phys.\ Rev.\  D {\bf 70}, 075015 (2004)
  [arXiv:hep-ph/0403144].

\bibitem{Ellis:2005jc}
  J.~R.~Ellis, V.~E.~Mayes and D.~V.~Nanopoulos,
  Phys.\ Rev.\  D {\bf 74}, 115003 (2006)
  [arXiv:astro-ph/0512303].

\bibitem{Kane:2004hm}
  G.~L.~Kane, P.~Kumar, J.~D.~Lykken and T.~T.~Wang,
  Phys.\ Rev.\  D {\bf 71}, 115017 (2005)
  [arXiv:hep-ph/0411125].

\bibitem{Font:2004cx}
  A.~Font and L.~E.~Ibanez,
  JHEP {\bf 0503}, 040 (2005)
  [arXiv:hep-th/0412150].

\bibitem{Djouadi:2002ze}
  A.~Djouadi, J.~L.~Kneur and G.~Moultaka,
  Comput.\ Phys.\ Commun.\  {\bf 176}, 426 (2007)
  [arXiv:hep-ph/0211331].

\bibitem{Belanger:2004yn}
  G.~Belanger, F.~Boudjema, A.~Pukhov and A.~Semenov,
  Comput.\ Phys.\ Commun.\  {\bf 174}, 577 (2006).

\bibitem{Aldazabal:2000cn}
  G.~Aldazabal, S.~Franco, L.~E.~Ibanez, R.~Rabadan and A.~M.~Uranga,
  JHEP {\bf 0102}, 047 (2001)
  [arXiv:hep-ph/0011132].

\bibitem{Chamoun:2003pf}
  N.~Chamoun, S.~Khalil and E.~Lashin,
  Phys.\ Rev.\  D {\bf 69}, 095011 (2004)
  [arXiv:hep-ph/0309169].

\bibitem{Higaki:2005ie}
  T.~Higaki, N.~Kitazawa, T.~Kobayashi and K.~j.~Takahashi,
  Phys.\ Rev.\  D {\bf 72}, 086003 (2005)
  [arXiv:hep-th/0504019].
  

\bibitem{Ross:2007az}
  G.~Ross and M.~Serna,
  arXiv:0704.1248 [hep-ph].


\bibitem{Fusaoka:1998vc}
  H.~Fusaoka and Y.~Koide,
  Phys.\ Rev.\  D {\bf 57}, 3986 (1998)
  [arXiv:hep-ph/9712201].

\bibitem{INPREP}
  Ching-Ming Chen, Tianjun Li, V. E. Mayes, James Maxin, D. V. Nanopoulos, in preparation. 
\end{thebibliography}
\end{document}